\documentclass[11pt,authoryear,review,3p,times]{elsarticle}
\usepackage[colorlinks=true,citecolor=blue,linkcolor=blue]{hyperref}
\usepackage[english]{babel}
\usepackage[utf8]{inputenc}
\usepackage[T1]{fontenc}
\usepackage{graphicx,array,bm,color,natbib,amssymb}
\usepackage{booktabs,caption}
\usepackage[flushleft]{threeparttable}
\usepackage{amsmath}
\usepackage{algorithm,algcompatible,amsmath}
\usepackage[noend]{algpseudocode}

\newdefinition{definition}{Definition}
\newproof{proof}{Proof}
\usepackage{subcaption}
\newcommand{\Tau}{\mathcal{T}}
\def\d{~\textnormal{d}}
\makeatletter
\def\BState{\State\hskip-\ALG@thistlm}
\makeatother

\journal{to the chosen journal}

\begin{document}
		
		\begin{frontmatter}
		\title{{ \LARGE Estimation of component reliability from superposed renewal processes with masked cause of failure by means of latent variables}}
     
		\author[label2]{\large Agatha Rodrigues \fnref{label3}\corref{aaa}}	
		\author{\large Pascal Kerschke \fnref{label4}}
		\author[label2]{\large Carlos Alberto de B. Pereira \fnref{label41}}
		\author{\large Heike  Trautmann \fnref{label4}}	
		\author{\large Carolin Wagner \fnref{label5}}	
		\author{\large Bernd Hellingrath \fnref{label5}}
		\author{\large Adriano Polpo\fnref{label6}}
	
		\address[label2]{Institute of Mathematics and Statistics, University of S\~ao Paulo,  S\~ao Paulo, SP, Brazil.}
    	\address[label3]{Department of Obstetrics and Gynecology, S\~ao Paulo University Medical School, S\~ao Paulo, SP, Brazil.}	
    	\address[label4]{Information Systems and Statistics, University of Münster, 48149 Münster, Germany.}    	
    	\address[label41]{Institute of Mathematics, Federal University of Mato Grosso do Sul, Campo Grande, MS, Brazil.}
    	\address[label5]{Information Systems and Supply Chain Management, University of Münster, 48149 Münster, Germany.}    
		\address[label6]{Western Australia University, Perth, Australia.}

		\cortext[aaa]{e-mail: agatha.srodrigues@gmail.com} 

\begin{abstract}
 
In a system, there are identical replaceable components working for a given task and a failed component is replaced by a functioning one in the corresponding position, which characterizes a repairable system. Assuming that a replaced component lifetime has the same lifetime distribution as the old one, a single component position can be represented by a renewal process and the multiple components positions for a single system form a superposed renewal process. When the interest consists in estimating the component lifetime distribution, there are a considerable amount of works that deal with estimation methods for this kind of problem. However, the information about the exact position of the replaced component is not available, that is, a masked cause of failure. In this work, we propose two methods, a Bayesian and a maximum likelihood function approaches, for estimating the failure time distribution of components in a repairable system with a masked cause of failure. As our proposed estimators consider latent variables, they yield better performance results compared to commonly used estimators from the literature. The proposed models are generic and straightforward for any probability distribution. Aside from point estimates, intervalar estimates are presented for both approaches. Using several simulations, the performances of the proposed methods are illustrated and their efficiency and applicability are shown based on the so-called cylinder problem.
 
\end{abstract}

\begin{keyword}
Bayesian paradigm, component lifetime, EM algorithm, Markov-Chain Monte-Carlo, maximum likelihood estimator, Metropolis within Gibbs algorithm, parametric estimation, repairable system, series system.
\end{keyword}
\end{frontmatter}

\section{Introduction}

A system of components is composed of components working for a given task. A failed component is replaced by an identical functioning one in the corresponding position, which characterizes a repairable system. Assuming that a replaced component lifetime has the same lifetime distribution as the old one, a single component position can be represented by a renewal process (RP). The multiple components positions form a superposed renewal process (SRP), that is, a single system can be seen as a SRP \citep{Rinne}. 
The objective is to estimate the failure time distribution of components that form the system and some approaches have been explored to analyze SRP data \citep{crowder1994statistical,Nelson2003,meeker2014statistical, Crow1990}.

However, there are situations in which the information about the exact position of the component replacement is not available, that is, there is the information that a component was replaced for a given system, but not information on which position the component was replaced.  
Cases like this are known as a masked cause of failure and have been considered in the literature in not repairable situations \citep{Miyakawa, SarhanEl, Muk2006, KuoYang2000, Fan2014, Wang2015, liu2017nonparametric, RodriguesEtAl2}.  
 
The scenario considered in this work is the following: a fleet of systems (sample) is observed. Within each system, there is a set of $m$ identical components and when a component fails, it is replaced by a functioning one in its position, which we will call socket. Although the number of failures $r$ within the interval $[0,\tau]$, $\tau$ is the end-of-observation time, can be observed for a given system, this information is unknown for the single sockets. 

\cite{zhang2017} propose a procedure for estimating the component lifetime distribution from a collection of SRPs with masked cause of failure by maximizing its likelihood function. The likelihood function is given by the sum of all possible data configurations, that is, all possible combinations in which the $r$ failures might occur across the $m$ sockets. However, the number of all possible data configurations increases exponentially with the number of failures, and for large numbers of $m$ and $r$, the computation of the maximum likelihood is too expensive. 
Thus, depending on the numbers of failures and components for each system in the fleet, the computational time is very costly and in some situations, it is not possible to compute. 
In this way, as the authors discuss, the method proposed by them is only applicable for dealing with a fleet of SRPs where each SRP only has a relatively small number of failures.

The aim of this work is to estimate the components' lifetime distribution involved in a collection of SRPs with masked cause of failure without restrictions about the numbers of components and failures. Our two methods -- a maximum likelihood and a Bayesian approach -- consider latent variables during the estimation process. 
 The contributions are as follows:
\begin{itemize}
	\item Under the maximum likelihood approach, we expect that considering latent variables and estimating the parameters via the Expectation Maximization (EM) algorithm \citep{RobertCasella} solves the limitation of the approach by \cite{zhang2017}, i.e, not  being able to compute the maximum likelihood estimator regardless of the number of failures and components. Besides, in situations in which the method of \cite{zhang2017} is useful, we expect that both methods yield similar performances, once they propose maximizing the likelihood function. 
    \item By proposing a Bayesian approach to solve the problem, we develop a useful method for incorporating expert knowledge and/or past experiences as a priori distribution, besides considering the statistical inference under the Bayesian paradigm.
\end{itemize}

Under the parametric approach, our proposed methods are generic and any probability distribution on positive support can be considered for the components' lifetime distributions. Aside from point estimates, interval estimates are discussed for both approaches.

The remainder of this manuscript is organized as follows. In section \ref{RP}, we describe the data structure. Sections \ref{MLA} and \ref{bayesian_approach} present the maximum likelihood and Bayesian approaches in more detail. Both methods are evaluated by means of simulation studies, in which they are compared with the method proposed by \cite{zhang2017}, in scenarios this last is possible, and the corresponding results are given in Section \ref{simulation}. Section \ref{cylinder_data} shows the applicability of the methodology in the cylinder dataset and Section \ref{final_remarks} concludes this work.

\section{Data structure} \label{RP}

Consider a system with $m$ components operating in $m$ sockets. Once a component fails, it is replaced by a new one in the same socket. In the following, we will define quantities for a single socket and hence omit the socket indices.

Let $Y_l$ denote the lifetime of the component before replacement $l$, for $l=1,2,\ldots$, under the assumption that the components' failure times are independent and identically distributed (i.i.d.).  
Besides, let $Z_{k}$ be a positive random variable that denotes the time of occurrence of the $k$-th failure in the socket. Thus, $Z_{k}=\sum_{l=1}^{k}Y_{l}$, $k\geq 1$,  and $\{Z_{k}\}$ is a renewal process (RP), that is, each socket in the system represents a RP. 


Once a system has $m$ independent sockets, each system-level set of failure times forms a superposed renewal process (SRP). Let $T_k$ be the $k$-th failure time of the system, in which $T_1=\min\{Y_{11},Y_{21},\ldots,Y_{m1}\}$ and $Y_{j1}$ denotes the first component failure time in the $j$-th socket, $j=1,\ldots,m$. 

Let $\bm{\Tau}=(t_1,t_2,\ldots,t_r,\tau)$ denote the observed event history of a single SRP with event times $t_1<t_2<\ldots<t_r$, and end-of-observation time $\tau$ with $\tau> t_r$. A data set will consist of $n$ independent SRPs corresponding to the $n$ systems in the fleet. 

In summary, the assumptions made here are: (a) the component distribution function is the same for all sockets and systems over time, (b) the failures within a socket are independent, (c) all sockets within one system have the same end-of-observation time $\tau$, and (d) the $n$ systems in the fleet are independent.

\section{Maximum likelihood approach} \label{MLA}

Under the assumption that the components' failure times are i.i.d., let $f(\cdot)=f(\cdot\mid \bm{\theta})$ and $R(\cdot)=R(\cdot \mid \bm{\theta})$ be the density and reliability functions of the component failure time, where $\bm{\theta}$ is a p-vector of unknown parameters. 

Consider a sample of $n$ systems. 
Let $\bm{t}_i=(t_{1i},t_{2i},\ldots,t_{r_ii})$ be the vector of observed $r_i$ failure times for the $i$-th system and $\tau_i$ the end-observation time, with $i=1,\ldots,n$, in which ${\bm \Tau}_i=(\bm{t}_i,\tau_i)$ is the observed data for the $i$-th system. Let $\bm{d}_i=(d_{1i},d_{2i},\ldots,d_{r_ii})$ the vector that indicates the cause of failure, in which $d_{ki}=j$, if component $j$ causes the $k$-th failure in the $i$-th system, for $j=1,\ldots,m$, $k=1,\ldots,r_i$ and $i=1,\ldots,n$.

Lets first assume that $\bm{d}_i$ is observed. As an example consider a system $i$ with $m = 16$ components for which $r_i = 3$ failures, $d_{i1} = d_{3i} = 1$ and $d_{2i} = 13$, were observed.  The likelihood contribution of this system is 
\begin{eqnarray}
f(t_{1i})f(t_{3i}-t_{1i})R(\tau_i-t_{3i})f(t_{2i})R(\tau_i-t_{2i})[R(\tau_i)]^{m-2}. \label{eq2}
\end{eqnarray}
Note that the likelihood contribution of system $i$ presents (\ref{eq2}) in a situation where $\bm{d}_i=(d_{1i},d_{2i},d_{3i})$ is known. In a masked cause of failure scenario, the actual failure position $\bm{d}_i$ of system $i$ are not observable. Hence, there are $V_i = m^{r_i} = 16^3 = 4,\!096$ possible configurations of likelihood contributions for this system, in which $V_i$ is the number of possible data configurations of system $i$ with $r_i$ failure times in $m$ components. The likelihood contribution of the $i$-th system is given by
\begin{eqnarray}
L_i=\sum_{v=1}^{V_i} L_{iv}, \nonumber
\end{eqnarray}
in which $L_{iv}$ is the likelihood contribution of the $v$-th configuration for system $i$. 
Considering that a fleet of $n$ independent systems is observed, the likelihood function for $\bm{\theta}$ is 
\begin{eqnarray}
L(\bm{\theta}\mid \bm{\Tau})=\prod_{i=1}^n\Bigg[\sum_{v=1}^{V_i} L_{iv}\Bigg], \label{vero_zhang}
\end{eqnarray}
where $\bm{\Tau}=(\bm{\Tau}_1,\ldots,\bm{\Tau}_n)$. \cite{zhang2017} propose the maximization of the likelihood function given in (\ref{vero_zhang}).

In the masked cause of failure scenario, $\bm{d}_i$ is a vector of latent variables. A suitable approach for estimating the parameter values, which maximize the likelihood function, is to consider an expectation-maximization (EM) algorithm. 
The latter is presented in the following subsection. 

\subsection{EM algorithm}  
The EM algorithm is an iterative method with Expectation (E) and Maximization (M) steps  \citep{dempster1977maximum}. The E-step evaluates the expectation of the full log-likelihood function and the M-step tries to find the parameter configuration, which maximizes the expectation found within the E-step. 

The augmented likelihood function (i.e., the likelihood function with latent variables) of $\bm{\theta}$ is given by
\begin{eqnarray} 
L(\bm{\theta}\mid  \bm{\Tau},\bm{d})=\prod_{i=1}^nL_{i}(\bm{\theta}\mid{\bm \Tau}_i,\bm{d}_i). \label{Full_vero}
\end{eqnarray}

The form of $L_{i}(\bm{\theta}\mid\bm{\Tau}_i,\bm{d}_i)$ depends on the number of failures $r_i$. For this reason, a general form is presented in the following. 

Given $\bm{d}_i$, let $\Gamma_i$ be the set of $v_i$ component indexes that cause at least one failure for system $i$. In a situation in which no failure is observed, $v_i=0$. Let $x_{ilk}$ the $k$-th failure time caused by the $l$-th element of $\Gamma_i$, with $l=1,\ldots,v_i$ and $k=1,\ldots,n_l$. As an example, for system $i$ with $r_i=3$ failures observed and $d_{1i}=d_{3i}=1$ and $d_{2i}=13$, we have $\Gamma_i=\{1,13\}$, $v_i=2$, $n_1=2$ and $n_2=1$, $x_{i11}=t_{1i}$, $x_{i12}=t_{3i}$ and $x_{i21}=t_{2i}$. Thus, $\sum_{l=1}^{v_i}n_l=r_i$.

The likelihood contribution of the $i$-th system can be written as
\begin{eqnarray} 
L_{i}(\bm{\theta}\mid{\bm\Tau}_i,\bm{d}_i)=\Bigg\{\prod_{l=1}^{v_i}\Bigg[\prod_{k=1}^{n_l}f(x_{ilk}-x_{il(k-1)})\Bigg]R(\tau_i-x_{iln_l})\Bigg\}^{1-\rm{I}(v_i=0)} R(\tau_i)^{m-v_i}, \nonumber  
\end{eqnarray}
with $x_{il0}=0$ and indicator function $\rm{I}(A)=1$, if $A$ is true.


 Let  $l_{i}(\bm{\theta}\mid{\bm\Tau}_i,\bm{d}_i)=\log L_{i}(\bm{\theta}\mid{\bm \Tau}_i,\bm{d}_i)$. Thus, the logarithm of the augmented likelihood in (\ref{Full_vero}) can be written as 
 \begin{eqnarray} 
&& l(\bm{\theta}\mid {\bm\Tau},\bm{d})=\sum_{i=1}^nl_{i}(\bm{\theta}\mid{\bm\Tau}_i,\bm{d}_i) \nonumber \\
                                 &=& \sum_{i=1}^n \Bigg\{\Big[1-\rm{I}(v_i=0)\Big]\Bigg[\sum_{l=1}^{v_i}\sum_{k=1}^{n_l}\log f(x_{ilk}-x_{il(k-1)}) + \sum_{l=1}^{v_i} \log R(\tau_i-x_{iln_l}) \Bigg]+ (m-v_i)\log R(\tau_i) \Bigg\} 
                                  \label{log_full_lik}
 \end{eqnarray}
 
 Let $\bm{\theta}_r$ be the value assumed by $\bm{\theta}$ in the $r$-th iteration of the algorithm. The $(r+1)$-th E-step consists of calculating the expectation of (\ref{log_full_lik}), that is, 
  \begin{eqnarray} 
 Q(\bm{\theta}\mid\bm{\theta}_r)={\rm E}\big[l(\bm{\theta}\mid {\bm\Tau},\bm{d})\mid{\bm\Tau};\bm{\theta}_r\big].  \label{expectation}
 \end{eqnarray}
Unfortunately, there exists no analytical expression of the expectation in (\ref{expectation}). Instead, it can be approximated by Monte-Carlo simulations. Consider that $L$ random samples $\bm{d}_{i}^{(1)},\ldots,\bm{d}_{i}^{(L)}$ are simulated based on $f(\bm{d}_{i}\mid{\bm\Tau})$, i.e., the density function of $\bm{d}$ conditional to ${\bm\Tau}$, $i=1,\ldots,n$ (see Subsection \ref{dist_cond_d}).  Thus, the E-step results in calculating 
\begin{eqnarray} 
  Q_m(\bm{\theta}\mid\bm{\theta}_r)= \frac{1}{L} \sum_{l = 1}^{L} l(\bm{\theta}\mid {\bm\Tau},\bm{d}^{(l)}) =\frac{1}{L}\sum_{l=1}^L\sum_{i=1}^nl_{i}\Big(\bm{\theta}\mid{\bm\Tau}_i,\bm{d}_{i}^{(l)}\Big). \label{expectation_MC}
 \end{eqnarray}
 The M-step maximizes (\ref{expectation_MC}) with respect to $\bm{\theta}$ resulting in $\bm{\theta}_{r+1}$. The optimization method considered within this work is the Nelder-Mead algorithm \citep{nelder1965simplex}. 
 The  E- and M-steps are alternated until the difference of estimates between two consecutive iteration values is less than $10^{-4}$. The estimate of $\bm{\theta}$, say  $\widehat{\bm{\theta}}$, is obtained when the convergence criterion is reached. In this work, we consider $L=1,\!000$. 

 Let $g(\bm{\theta})$ be a function of $\bm{\theta}$. Due to the invariance property of the maximum likelihood estimator (MLE), the MLE of $g(\bm{\theta})$ is $g(\widehat{\bm{\theta}})$. For instance, if the Weibull distribution with parameters $\beta>0$ (shape) and $\eta>0$ (scale) is assumed for components' failure times, in wich  $\bm{\theta}= (\beta,\eta)$, the expected time of the component's lifetime is $\rm{E}(Y)=g(\bm{\theta})=\eta\Gamma(1+(1/\beta))$ and its MLE is $g(\widehat{\bm{\theta}})=\widehat{\eta}\Gamma(1+(1/\widehat{\beta}))$, in which $\widehat{\beta}$ and $\widehat{\eta}$ are MLE of $\beta$ and $\eta$, respectively \citep{casella2002statistical}. In an analogous way, the MLE for the component reliability function is $\widehat{R}(y)=\exp\Big[-(y/\widehat{\eta})^{\widehat{\beta}}\Big]$, for $y>0$.
  
\subsubsection{Conditional distribution of d given $\Tau$}  \label{dist_cond_d}
 
  
 For a fixed $i$, $f(\bm{d}_i\mid{\bm\Tau}_i )$ can be written as
  \begin{eqnarray} 
f(\bm{d}_i\mid{\bm\Tau}_i)&=& f(d_{1i},d_{2i},\ldots,d_{r_ii}\mid{\bm\Tau}_i) \nonumber \\ &=&f(d_{r_ii}\mid{\bm\Tau}_i,d_{(r_{i}-1)i},d_{(r_{i}-2)i},\ldots,d_{2i},d_{1i})f(d_{(r_{i}-1)i}\mid{\bm\Tau}_i,d_{(r_{i}-2)i},\ldots,d_{2i},d_{1i})\ldots f(d_{2i}\mid{\bm\Tau}_i,d_{1i}) \nonumber \\
&& ~~~~~ \times f(d_{1i}\mid{\bm\Tau}_i).\nonumber
 \end{eqnarray}

As an example, consider $r_i=3$ and ${\bm\Tau}_i=(t_{1i},t_{2i},t_{3i},\tau_i)$. Thus,
 \begin{eqnarray} 
f(\bm{d}_i\mid{\bm\Tau}_i)= f(d_{1i},d_{2i},d_{3i}\mid{\bm\Tau}_i) =f(d_{3i}\mid{\bm\Tau}_i,d_{2i},d_{1i})f(d_{2i}\mid{\bm\Tau}_i,d_{1i})f(d_{1i}\mid{\bm\Tau}_i).\nonumber
\end{eqnarray}
Under i.i.d assumption, the distribution of $d_{1i}=j\mid {\bm\Tau}_i$ follows a Multinomial distribution, that is, $Multin(1,\bm{p}_{1i})$, with $\bm{p}_{1i}=(p_{11i},\ldots,p_{1mi})$ and $p_{1ji}=1/m$, $j=1,\ldots,m$. Note that in this special case, the multinomial distribution equals a discrete uniform distribution.

Similarly, the distribution of $d_{2i}\mid({\bm\Tau}_i,d_{1i}=j)$ can be described as follows:  
	 \begin{eqnarray} 
	f(d_{2i}\mid{\bm\Tau}_i,d_{1i}=j)\propto [f(t_{2i}-t_{1i})]^{\rm{I}(d_{2i}=j)}\prod_{l=1; l\neq j}^m[f(t_{2i})]^{\rm{I}(d_{2i}=l)},\nonumber
	\end{eqnarray}
     that is, $d_{2i}\mid(\bm{t}_i,d_{1i}=j)$ follows $Multin(1,\bm{p}_{2i})$, in which $\bm{p}_{2i}=(p_{21i},\ldots,p_{2mi})$, $p_{2ji}=f(t_{2i}-t_{1i})/C$ and $p_{2li}=f(t_{2i})/C$, $l=1,\ldots,m$ and $l\neq j$, with $C=f(t_{2i}-t_{1i}) +(m-1)f(t_{2i})$.
     
For the conditional distribution of $d_{3i}$, one has to consider the following two cases:  
    \begin{itemize}
    	\item Distribution of $d_{3i}\mid({\bm\Tau}_i,d_{1i}=j,d_{2i}=j)$:
         \begin{eqnarray} 
    	   f(d_{3i}\mid{\bm\Tau}_i,d_{1i}=j,d_{2i}=j)\propto [f(t_{3i}-t_{2i})]^{\rm{I}(d_{3i}=j)}\prod_{l=1; l\neq j}^m[f(t_{3i})]^{\rm{I}(d_{3i}=l)},\nonumber
    	\end{eqnarray}
    	that is, $d_{3i}\mid({\bm\Tau}_i,d_{1i}=j,d_{2i}=j)$ follows $Multin(1,\bm{p}_{3i})$, in which  $\bm{p}_{3i}=(p_{31i},\ldots,p_{3mi})$, $p_{3ji}=f(t_{3i}-t_{2i})/C$ and $p_{3li}=f(t_{3i})/C$, $l=1,\ldots,m$ and $l\neq j$, with $C=f(t_{3i}-t_{2i}) +(m-1)f(t_{3i})$.
    	\item Distribution of $d_{3i}\mid({\bm\Tau}_i,d_{1i}=j,d_{2i}=q)$, with $q\neq j$:
    	\begin{eqnarray} 
    	f(d_{3i}\mid{\bm\Tau}_i,d_{1i}=j,d_{2i}=q)\propto [f(t_{3i}-t_{1i})]^{\rm{I}(d_{3i}=j)}[f(t_{3i}-t_{2i})]^{\rm{I}(d_{3i}=q)}\prod_{l=1; l\neq j,q}^m[f(t_{3i})]^{\rm{I}(d_{3i}=l)},\nonumber
    	\end{eqnarray}
    	that is, $d_{3i}\mid({\bm\Tau}_i,d_{1i}=j,d_{2i}=q)$ follows $Multin(1,\bm{p}_{3i})$, in which  $\bm{p}_{3i}=(p_{31i},\ldots,p_{3mi})$, $p_{3ji}=f(t_{3i}-t_{1i})/C$, $p_{3qi}=f(t_{3i}-t_{2i})/C$ and $p_{3li}=f(t_{3i})/C$, $l=1,\ldots,m$ and $l\neq j,q$, with $C=f(t_{3i}-t_{1i})+f(t_{3i}-t_{2i}) +(m-2)f(t_{3i})$.
    \end{itemize}

\subsection{Asymptotic Distribution}


The asymptotic distribution of the maximum likelihood estimator $\widehat{\bm{\theta}}$ can be approximated by a multivariate normal distribution with mean $\bm{\theta}$ and variance-covariance matrix 
$I_{\bm{\theta}}(\bm{\theta})^{-1}$, where $I_{\bm{\theta}}(\bm{\theta})$ is the observed information matrix for $\bm{\theta}$. As demonstrated by \citet{louis1982finding}, $I_{\bm{\theta}}(\widehat{\bm{\theta}})$ is the sum of 
\begin{eqnarray}
	I_1(\bm{\theta}\mid\widehat{\bm{\theta}})=-\frac{\partial^2}{\partial\bm{\theta}\partial\bm{\theta}^\top}Q(\bm{\theta}\mid\widehat{\bm{\theta}}) \quad \mbox{and} \quad I_2(\bm{\theta}\mid\widehat{\bm{\theta}})=-{\rm Var}\Bigg\{\frac{\partial}{\partial\bm{\theta}}l(\bm{\theta}\mid {\bm\Tau},\bm{d})\Big|{\bm\Tau};\widehat{\bm{\theta}}\Bigg\}. \nonumber
\end{eqnarray}
The matrix $I_1(\bm{\theta}\mid \widehat{\bm{\theta}})$ can be estimated by
\begin{eqnarray}
	-\frac{\partial^2}{\partial\bm{\theta}\partial\bm{\theta}^\top}Q_m(\bm{\theta}\mid\widehat{\bm{\theta}})=-\frac{1}{L}\sum_{i=1}^n\sum_{l=1}^L\frac{\partial^2}{\partial\bm{\theta}\partial\bm{\theta}^\top}l_{i}\Big(\bm{\theta}\mid{\bm\Tau}_i,\bm{d}_{i}^{(l)}\Big)\Bigg|_{\bm{\theta}=\widehat{\bm{\theta}}}, \nonumber 
\end{eqnarray}
where $\bm{d}_{i}^{(l)}$, with $l=1,\ldots,L$, being a random sample from the distribution of $f(\bm{d}_{i}\mid{\bm\Tau}_i)$ for the $i$-th system. 

An estimate of $I_2(\bm{\theta}\mid\widehat{\bm{\theta}})$ results from the sum of 
\begin{eqnarray}
	\sum_{i=1}^n\Bigg\{\frac{1}{L}\sum_{l=1}^L\frac{\partial}{\partial\bm{\theta}}l_{i}\Big(\bm{\theta}\mid{\bm\Tau}_i,\bm{d}_{i}^{(l)}\Big)\Bigg|_{\bm{\theta}=\widehat{\bm{\theta}}}\Bigg\}\Bigg\{\frac{1}{L}\sum_{l=1}^L\frac{\partial}{\partial\bm{\theta}}l_{i}\Big(\bm{\theta}\mid{\bm\Tau}_i,\bm{d}_{i}^{(l)}\Big)\Bigg|_{\bm{\theta}=\widehat{\bm{\theta}}}\Bigg\}^\top \nonumber 
\end{eqnarray}
and
\begin{eqnarray}
	-\frac{1}{L}\sum_{i=1}^n\sum_{l=1}^L\Bigg\{\frac{\partial}{\partial\bm{\theta}}l_{i}\Big(\bm{\theta}\mid{\bm\Tau}_i,\bm{d}_{i}^{(l)}\Big)\Bigg\}\Bigg\{\frac{\partial}{\partial\bm{\theta}}l_{i}\Big(\bm{\theta}\mid{\bm\Tau}_i,\bm{d}_{i}^{(l)}\Big)\Bigg\}^\top\Bigg|_{\bm{\theta}=\widehat{\bm{\theta}}}. \nonumber
\end{eqnarray}

Detailed information on the development of $I_{\bm{\theta}}(\widehat{\bm{\theta}})^{-1}$ if one assumes Weibull distribution with parameters $\beta$ (shape) and $\eta$ (scale) is given in the appendix.

Thus, an asymptotic $\gamma\%$ confidence interval for $\bm{\theta}$ $(CI\gamma\%)$ is given by
\begin{eqnarray}
CI\gamma\%=\Big(\widehat{\bm{\theta}}-z_{(1-\gamma/2)}\sqrt{(I_{11},\ldots,I_{pp})}; ~~~ \widehat{\bm{\theta}}+z_{(1-\gamma/2)}\sqrt{(I_{11},\ldots,I_{pp})}\Big), \nonumber
\end{eqnarray}
in which $I_{jj}$ denotes the $j$th element of the main diagonal of $I_{\bm{\theta}}(\widehat{\bm{\theta}})^{-1}$.

Confidence intervals for functions of ${\bm {\theta}}$ can be obtained by the delta method \citep{casella2002statistical}.

\subsection{Model selection criteria}
One can consider some discrimination criteria to select the model based on the maximized log-likelihood function. They are: AIC (Akaike Information Criterion), AICc (Corrected Akaike Information Criterion), BIC (Bayesian Information Criterion), HQIC (Hannan-Quinn Information Criterion) and CAIC (Consistent Akaike Information Criterion), which are computed,
respectively, by $AIC = 2p-2l$, $AICc = AIC + 2p(p + 1)/(n -p - 1)$, $BIC = p \log n-2l $, $HQIC =  2p \log(\log n)-2l$ and $CAIC = p(\log n + 1)-2l$, where $p$ is the number of parameters of the fitted model, $n$ is the sample size and $l$ is the maximized log-likelihood function value, obtained by evaluating (\ref{expectation_MC}) in the last iteration of EM algorithm estimates. 

Given a set of candidate models, the preferred model is the one which provides the minimum criteria values.

\section{Bayesian Approach}\label{bayesian_approach}

In the Bayesian approach, the latent variable vector $\bm{d}$ is faced as parameter vector. Thus, the posterior distribution of $(\bm{\theta},\bm{d})$ can be written as
 \begin{eqnarray} 
 \pi(\bm{\theta},\bm{d}  \mid {\bm\Tau}) & ~~ \propto &  ~~ \pi(\bm{\theta}, \bm{d})  L(\bm{\theta}, \bm{d} \mid {\bm\Tau}),    \label{posteriori}
 \end{eqnarray}
 where $L(\bm{\theta}, \bm{d} \mid {\bm\Tau})$ has the same form as (\ref{Full_vero}) in which $\bm{d}$ now is faced as parameter and $\pi(\bm{\theta}, \bm{d})$ is the prior distribution of $(\bm{\theta}, \bm{d})$. 
 
 In real-world settings, it is possible that the prior distributions can be influenced by expert knowledge and/or past experiences on the functioning of the components. In this work, no prior information about the functioning of the components is available, which is the reason for the choice of non-informative prior distributions, besides of the assumption that the parameters are independent a prior. 
 
 Given the posterior density in Equation (\ref{posteriori}) does not have a closed form, statistical inferences about the parameters can rely on Markov-Chain Monte-Carlo (MCMC) simulations. Here, we consider the Metropolis within Gibbs algorithm \citep{Tierney} once it is possible to sample some of the parameters directly from the conditional distribution; however, this is not possible for other parameters. The algorithm works in the steps presented in Algorithm \ref{metro_with_gibbs}. 
 
 
\begin{algorithm}[!h]
	\caption{The Metropolis within Gibbs algorithm.}\label{metro_with_gibbs}
	\begin{algorithmic}[1]
		\State Assign initial values $\bm{\theta}^{(0)}$ for   $\bm{\theta}$ and set $b=1$.
		\State Draw $\bm{d}_{i}^{(b)}$ from $\pi(\bm{d}_i\mid{\bm\Tau}_i,\bm{\theta})$  from 
		\begin{eqnarray}
		\pi(\bm{d}_i\mid{\bm\Tau}_i,\bm{\theta})&=& \pi(d_{1i},d_{2i},\ldots,d_{r_ii}\mid{\bm\Tau}_i,\bm{\theta}) \nonumber \\ &=&\pi(d_{r_ii}\mid{\bm\Tau}_i,\bm{\theta},d_{(r_{i}-1)i},d_{(r_{i}-2)i},\ldots,d_{2i},d_{1i})\pi(d_{(r_{i}-1)i}\mid{\bm\Tau}_i,\bm{\theta},d_{(r_{i}-2)i},\ldots,d_{2i},d_{1i})\times \nonumber \\ &&\ldots\times\pi(d_{2i}\mid{\bm\Tau}_i,\bm{\theta},d_{1i})\pi(d_{1i}\mid{\bm\Tau}_i,\bm{\theta}),\nonumber
		\end{eqnarray}
		in an analogous way presented in Subsection \ref{dist_cond_d}, for $i=1,\ldots,n$, and $ \bm{d}^{(b)}=(\bm{d}_{1}^{(b)},\ldots,\bm{d}_{n}^{(b)})$.
		\State Draw $\bm{\theta}^{(b)}$ from 
		\begin{eqnarray}
			\pi(\bm{\theta}\mid {\bm\Tau}, \bm{d}^{(b)}) \propto \pi(\bm{\theta})\prod_{i=1}^n\Bigg\{ \Bigg[\prod_{l=1}^{v_i}\Bigg(\prod_{k=1}^{n_l}f(x_{ilk}-x_{il(k-1)})\Bigg)R(\tau_i-x_{iln_l})\Bigg]^{1-\rm{I}(v_i=0)} R(\tau_i)^{m-v_i}\Bigg\}, \nonumber 
		\end{eqnarray}
	   through Metropolis-Hastings algorithm \citep{RobertCasella}. 
		\State Set $b=b+1$ and repeat steps $2)$ and $3)$ until $b=B$, where $B$ is the predefined number of simulated samples of $(\bm{\theta},\bm{d})$.		
	\end{algorithmic}
\end{algorithm}
  
Discarding burn-in (i.e., the first generated values are discarded to eliminate the effect of the assigned initial values for parameters) and jump samples (i.e., gaps between the generated values in order to avoid correlation problems), a sample of size $n_p$ from the joint posterior distribution of $(\bm{\theta},\bm{d})$ is obtained. The sample from the posterior distribution can be expressed as $({\bm \theta}_{1},{\bm \theta}_{2},\ldots,{\bm \theta}_{n_p})$. Posterior quantities of $\bm{\theta}$ can be easily obtained \citep{RobertCasella}. For instance, the posterior mean of $\bm{\theta}$ can be approximated by 
 \begin{eqnarray}
 \frac{1}{n_p}\sum_{k=1}^{n_p}{\bm{\theta}_{k}}. \nonumber 
 \end{eqnarray}
The sample from the posterior distribution of $g(\bm{\theta})$ can be expressed as $(g({\bm \theta}_{1}),g({\bm \theta}_{2}),\ldots,g({\bm \theta}_{n_p}))$ and posterior quantities of $g(\bm{\theta})$ can be obtained. For instance, the posterior mean of the reliability function can be approximated by 
 \begin{eqnarray}
 \frac{1}{n_p}\sum_{k=1}^{n_p}{R(t \mid \bm{\theta}_{k})},~~ t > 0. \nonumber 
 \end{eqnarray}


The proposed approach is generic and straightforward for any probability distribution. Thus, it may be of interest to consider a model selection criterion. Below a criterion based on the conditional predictive ordinates is presented.

\subsection{Conditional predictive ordinate} 
A criterion for model selection that can be considered is based on the conditional predictive ordinates (CPO). For the $i$-th system, the conditional predictive ordinate (CPO) can be expressed as
 \begin{eqnarray} 
CPO_i=f(\bm{\Tau}_i\mid{\bm\Tau}_{-i})&&=\sum_{\bm{d}}\int{f(\bm{\Tau}_i\mid\bm{\theta},\bm{d})\pi(\bm{\theta},\bm{d}  \mid {\bm\Tau}_{-i})\partial{\bm{\theta}}} \nonumber \\
                                     &&= \Bigg\{\sum_{\bm{d}}\int{\frac{\pi(\bm{\theta},\bm{d}  \mid {\bm\Tau})}{f(\bm{\Tau}_i\mid\bm{\theta},\bm{d})}\partial{\bm{\theta}}}\Bigg\}^{-1} \nonumber \\
                                     && \approx  \Bigg\{\frac{1}{n_p}\sum_{k=1}^{n_p}\frac{1}{f(\bm{\Tau}_i\mid\bm{\theta}_k,\bm{d}_k)}\Bigg\}^{-1}, \nonumber 
\end{eqnarray}
in which $\bm{\Tau}_{-i}=(\bm{\Tau}_1,\ldots,\bm{\Tau}_{i-1},\bm{\Tau}_{i+1},\ldots,\bm{\Tau}_n)$ and $(\bm{\theta}_k,\bm{d}_k)$, for $k=1,\ldots,n_p$, represent a sample from the posterior distribution of $(\bm{\theta},\bm{d})$. 

High values of $CPO_i$ indicate that the model is capable of describing the $i$-th observation adequately \citep{gilks1995markov}. The LPML (log pseudo marginal likelihood) measure is the sum of the logarithms of the CPO of all the observations, that is, $LPML = \sum_{i=1}^n\log\Big(CPO_i\Big)$ and the higher the $LPML$ value is, the better the model fit.

 \section{Model evaluation by means of a simulation study} \label{simulation}
 
This section presents the results from simulation studies to evaluate the performance of the estimation methods described above in regards to the estimation quality. In scenarios the method of \cite{zhang2017} works, we compare its performance with those of the proposed methods. 

Thus, the following estimation methods were fitted: Bayesian approach (BA), maximum likelihood estimator via EM algorithm (EM-ML) and maximum likelihood estimator obtained by \citet{zhang2017} (Z-ML). The Z-ML estimates were obtained by means of the R-package \citep{softwareR} \texttt{SRPML} \citep{SRPML}.

The steps for generating the data of each simulated example, with $m$ being the number of sockets and $n$ the sample size, are presented in Algorithm \ref{data_generation}. The mean (7) and variance (4) values of component failure time distribution are based on cylinder application data (Section \ref{cylinder_data}). 

\begin{algorithm}[!h]
	\caption{Data generation.}\label{data_generation}
	\begin{algorithmic}[1]
		\For{each system unit $i=1,\ldots,n$}		
       		\State Draw $\tau_i$ from a Weibull distribution with mean $m_c$ and variance $0.05$. 
		    \State Draw $Y_{11i},Y_{21i},\ldots,Y_{m1i}$ from a Weibull distribution with mean $7$ and variance $4$, where $Y_{j1i}$ is the first component failure time in the $j$-th socket, for $j=1,\ldots,m$.
	     	\State Let $T_{1i}=\min\Big\{Y_{11i},Y_{21i},\ldots,Y_{m1i}\Big\}$.  
	     	\If  {$T_{1i} \geq \tau_i$}
	     	 \State stop simulation process and $r_i=0$.
		    \Else
		      \State Let $Y_{l1i}=\min\Big\{Y_{11i},Y_{21i},\ldots,Y_{m1i}\Big\}$, then $t_{1i}=Y_{l1i}$. 
		      \State Draw $Y_{l2i}$ from Weibull distribution with mean $7$ and variance $4$ conditional to $Y_{l2i}> t_{1i}$, where $Y_{l2i}$ is the second component failure time in the $l$-th socket, once the first failure occurred in the $l$-th socket. 
		      \State Let $T_{2i}=\min\Big\{Y_{11i},Y_{21i},\ldots,Y_{l2i},\ldots,Y_{m1i}\Big\}$. 
		      \If {$T_{2i} \geq \tau_i$} 
		       \State stop simulation process and $r_i=1$.
		      \Else
		       \State repeats steps $8$ to $10$ until $T_{r_i} <\tau_i< T_{(r_i+1)}$.
		    \EndIf
	       \EndIf
	   \EndFor
      \State The dataset is ${\bm \Tau}_i=\{t_{1i},t_{2i},\ldots,t_{r_ii},\tau_i\}$, for $i=1,\ldots,n$.
	\end{algorithmic}
\end{algorithm}

In this section, the Weibull distribution with parameters $\beta>0$ (shape) and $\eta>0$ (scale) is assumed for components' failure times, in wich  $\bm{\theta}= (\beta,\eta)$. For BA, the priors of Weibull parameters are considered to be independent gamma distributed with mean $1$ and variance $100$. Besides,  $d_{li}$ follows $Multin(1,\bm{p}_{li})$, where $\bm{p}_{li}=(p_{l1i},\ldots,p_{lmi})$ and $p_{lji}=1/m$, with $j=1,\ldots,m$. 

To obtain posterior quantities, we used an MCMC procedure to generate a sample from the posterior distribution of the parameters. We generated $20,\!000$ samples from the posterior distribution of each parameter. The first $10,\!000$ of these samples were discarded as burn-in samples. A jump of size $10$ was chosen to reduce correlation effects between the samples. As a result, the final sample size of the parameters generated from the posterior distribution was $1,\!000$. The chains' convergence was monitored in all simulation scenarios for good convergence results to be obtained. 

The mean absolute error (MAE) from each estimator to the true reliability of each method is considered as performance measure. $R(t)$ and $\widehat{R}(t)$ are the true reliability function and the estimate, respectively. Hence, the MAE is evaluated by $\frac{1}{l}\sum_{\ell=1}^{l} \mid \widehat{R}(g_{\ell})-R(g_{\ell}) \mid$, where $\{g_1, \ldots, g_{\ell}, \ldots, g_l \}$ is a grid in the space of failure times.  

First, we conducted two simulated examples, presented in the following. Second, scenarios with different sample sizes, number of sockets and censor mean time are considered. 

\subsection{Simulated examples}
 
We conducted two simulated examples considering $n=100$,  $m=16$ and $m_c=4$ (Example 1) or $m_c=8$ (Example 2), in which $m_c$ represents the mean of censor distribution, considered in step 2 in Algorithm \ref{data_generation}. It is worth noting that the expected number of failures with $m_c=8$ is larger than with $m_c=4$. 
 
For the Bayesian approach, the Gelman–Rubin convergence diagnostic statistics \citep{gelman1992inference} for parameters $\beta$ and $\eta$ are $1.0011$ and $1.0004$, respectively, in Example 1 and they are $1.0002$ and $1.0027$ in Example 2. 
 The measures are close to $1$, which suggests that convergence chains have been reached. 

\begin{figure}[!h]\centering
	\begin{minipage}[b]{0.47\linewidth}
		\includegraphics[width=\linewidth]{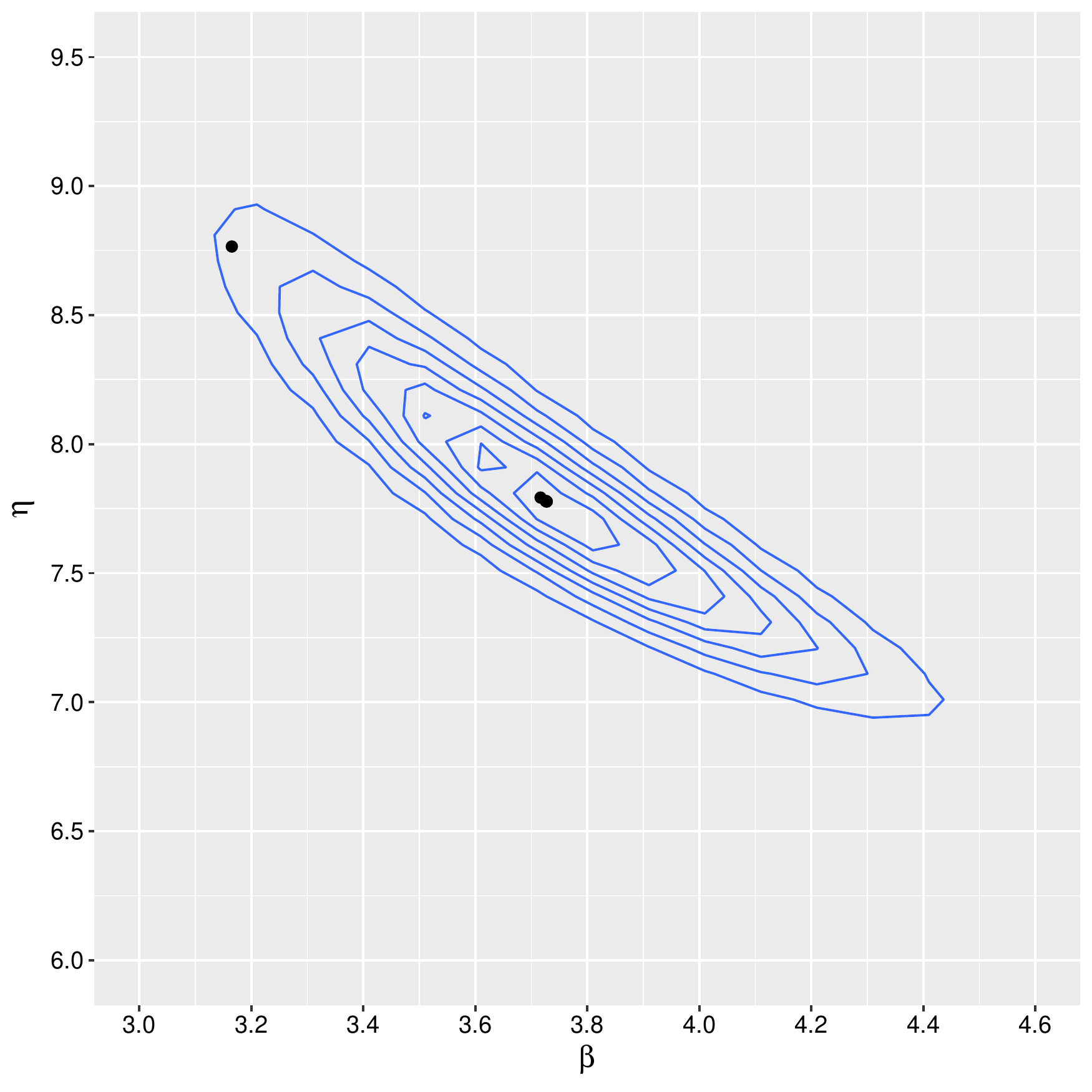}
		\subcaption{ $\mbox{   }$ Example 1}
		\label{Contour_SimulatedExample_mc4}
	\end{minipage} 
	\begin{minipage}[b]{0.47\linewidth}
		\includegraphics[width=\linewidth]{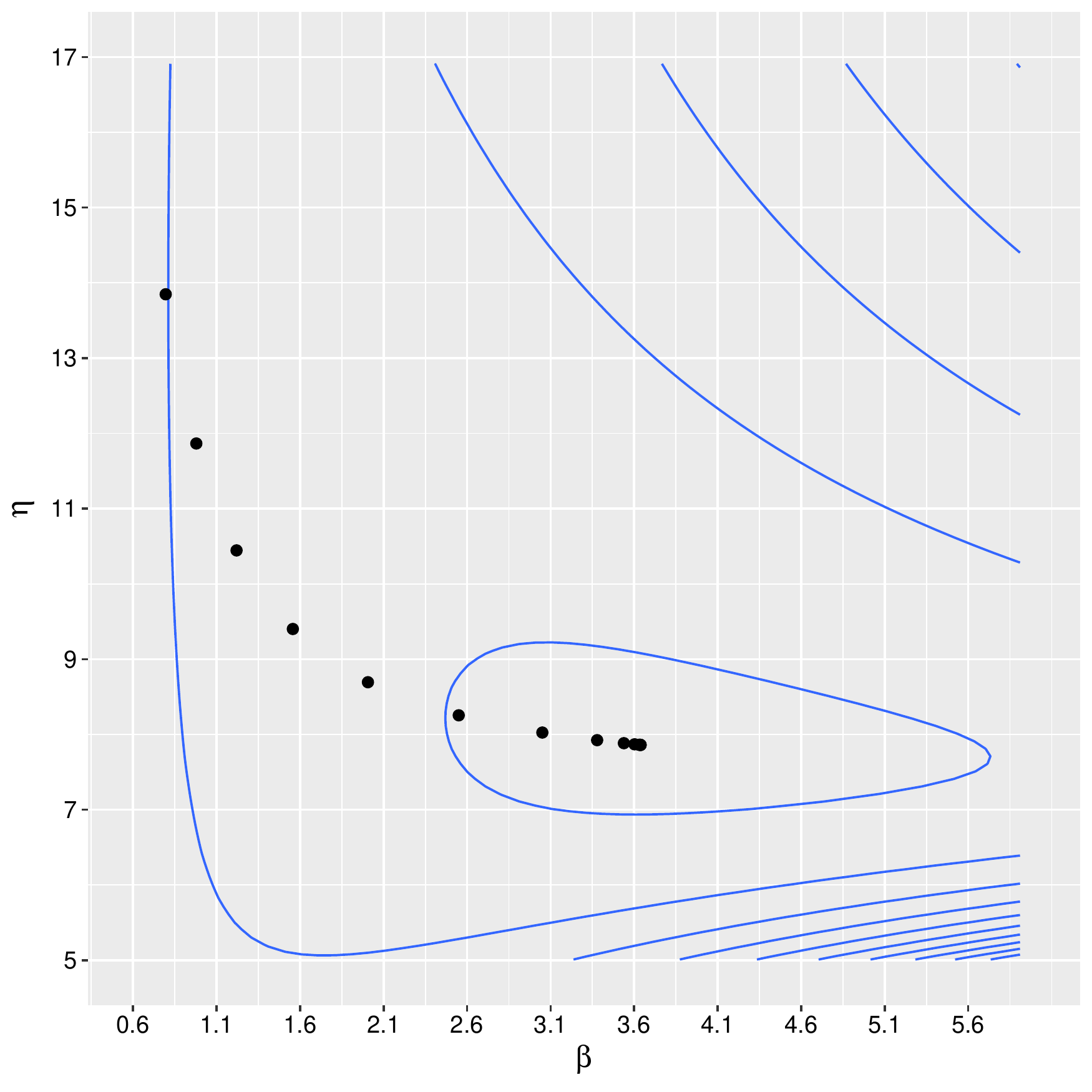}
		\subcaption{ $\mbox{   }$ Example 2}
		\label{Contour_SimulatedExample_mc8}
	\end{minipage}
	\caption{Contour plots of the log-likelihood function and EM algorithm iteration values (dots) for Example 1 (with $m_c=4$) and for Example 2 (with $m_c=8$), in which $m_c$ indicates the expected end-of-observation time.}
	\label{Contour_SimulatedExamples}
\end{figure}

For EM-ML, $8$ and $17$ EM iterations have been executed for Examples 1 and 2, respectively, and the corresponding values are listed in $\mbox{Table}~ \ref{iterations_SimulatedExamples}$. For both examples, the initial values for $(\beta,\eta)$ are $(1,1)$. After the first iteration it was $(1.206, 32.335)$ for Example 1, after the second one it was $(3.165, 8.766)$ and then reached the covergence region. For Example 2, it took about eight iterations to reach the covergence region. $\mbox{Figure}~\ref{Contour_SimulatedExamples}$ presents contour plots of the log-likelihood function, as well as the iteration values from the second to the eighth iteration for Example 1 and from third to $17$-th iteration for Example 2. The convergence was obtained fast for both examples.

The Weibull parameter estimates obtained by BA, EM-ML and Z-ML are presented in $\mbox{Table} ~ \ref{tab_estParams_ExSim}$. Note that the Z-ML estimation is not presented for Example 2, because those values could not be computed due to the high number of components and failures. The details about limitations of this method in situation of high numbers of failures and components are given in \citet{zhang2017}. 

\begin{figure}[!h]\centering
	\begin{minipage}[b]{0.47\linewidth}
		\includegraphics[width=\linewidth]{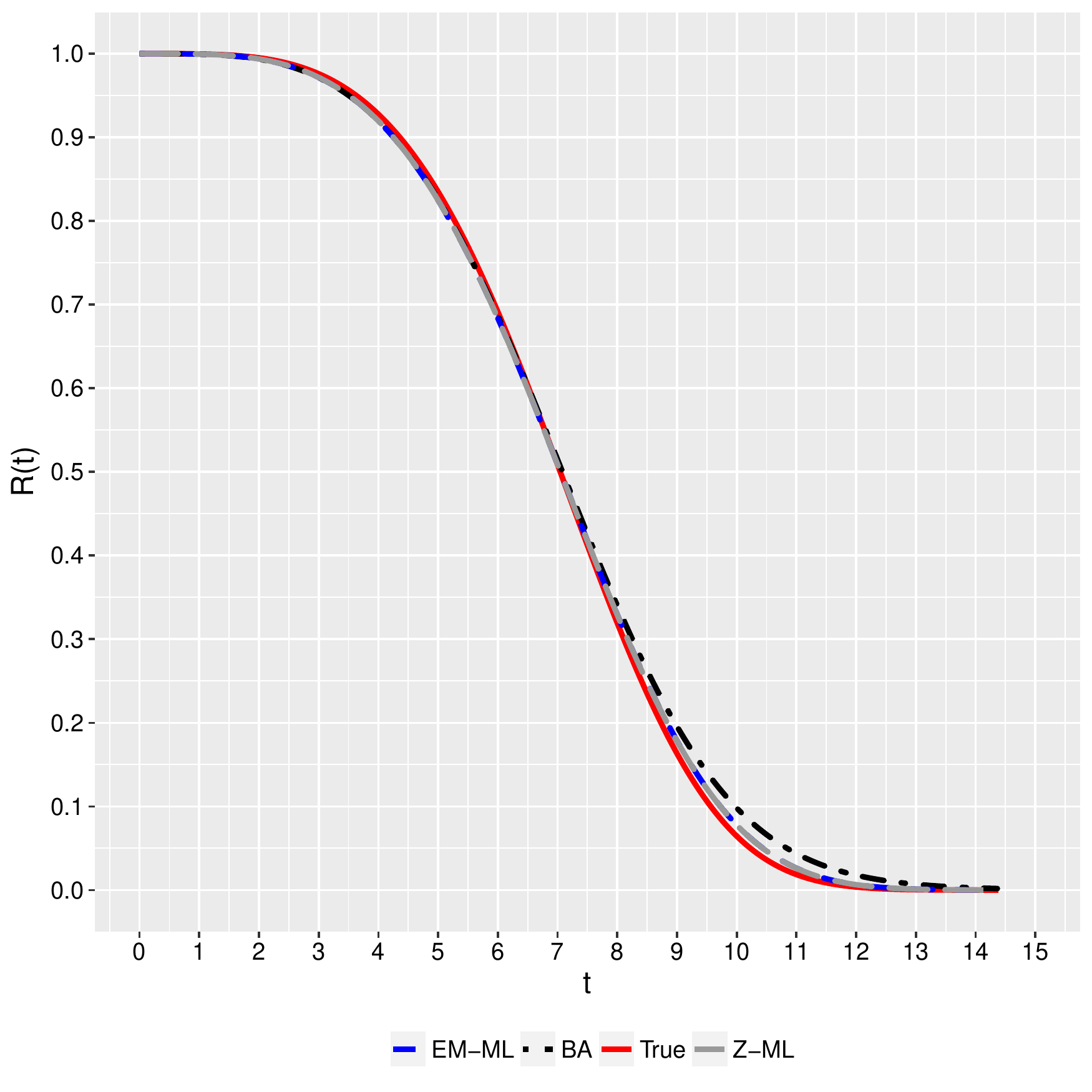}
		\subcaption{Example 1}
	\end{minipage} 
	\begin{minipage}[b]{0.47\linewidth}
		\includegraphics[width=\linewidth]{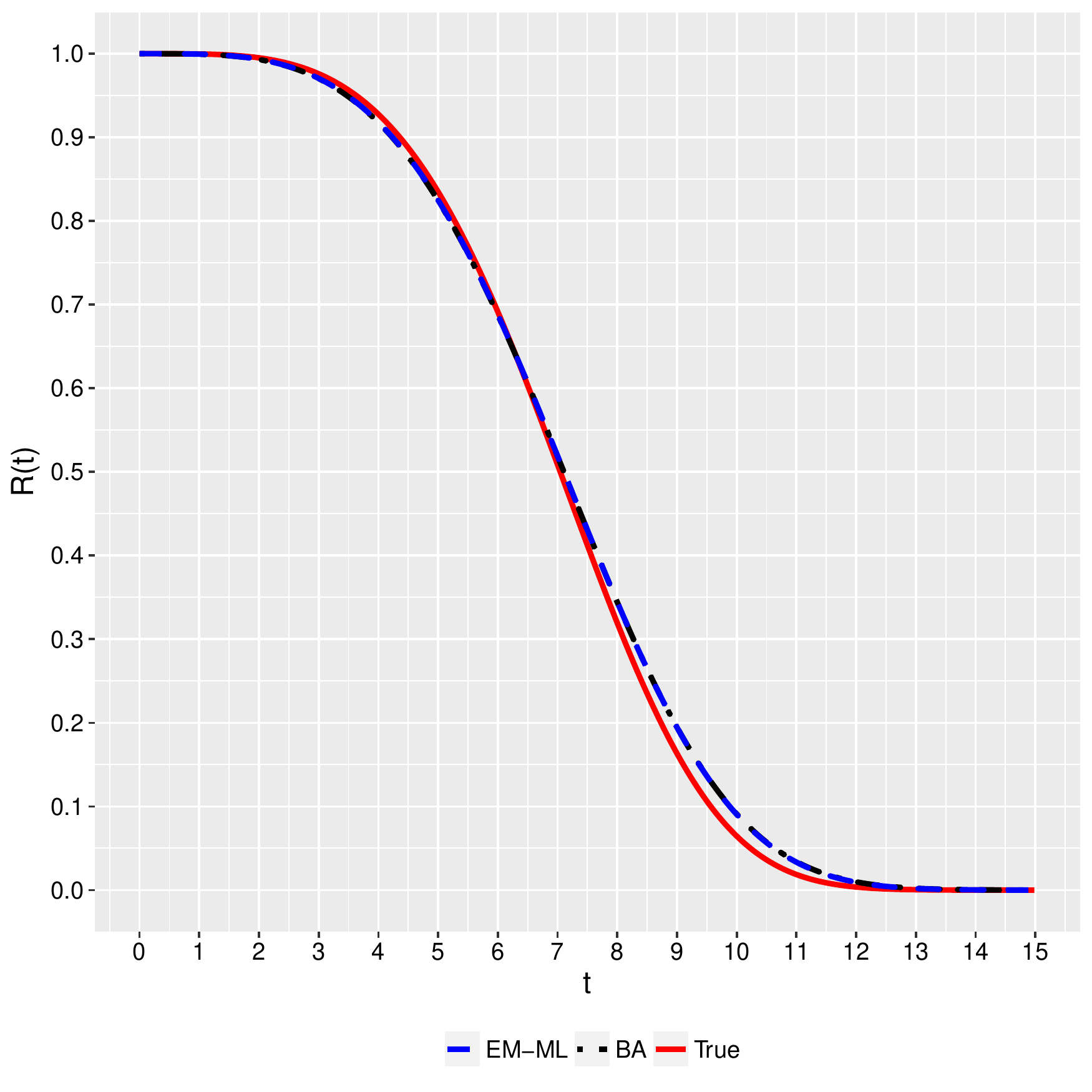}
		\subcaption{ Example 2}
	\end{minipage}
	\caption{Component reliability function estimation through the Bayesian approach (BA), the EM maximum likelihood method (EM-ML) and the maximum likelihood approach from {\it Zhang et al.} (Z-ML) for two scenarios of simulated examples, besides the generating curve (true). There is no Z-ML curve in Example 2 because it could not be computed due to the high number of components and failures.}
	\label{Relia_exampleSimulated}
\end{figure}

The estimates for the component reliability function obtained by BA, EM-ML, Z-ML, as well as the true reliability function, are presented in $\mbox{Figure}~ \ref{Relia_exampleSimulated}$. $\mbox{Table}~\ref{EMA_simulated_example}$ lists the MAE values, in which maximum likelihood approaches (EM-ML and Z-ML) present lower MAE values for Example 1, whereas BA and EM-ML present similar MAE values for Example 2. 

\begin{table}[h]
	\centering
	\caption{EM algorithm iteration values of Weibull parameters for two simulated examples.}
	\begin{tabular}{c|cc|cc}
		\hline
		& \multicolumn{2}{c|}{Example 1} & \multicolumn{2}{c}{Example 2} \\
		\cmidrule{2-5}    \multicolumn{1}{c|}{Iterations} & $\beta$ & $\eta$ & $\beta$ & $\eta$ \\
		\hline
		Initial value & 1.000 & 1.000 & 1.000 & 1.000 \\
		1     & 1.206 & 32.335 & 0.400 & 31.479 \\
		2     & 3.165 & 8.766 & 0.627 & 17.346 \\
		3     & 3.716 & 7.793 & 0.796 & 13.848 \\
		4     & 3.726 & 7.780 & 0.979 & 11.866 \\
		5     & 3.727 & 7.778 & 1.220 & 10.445 \\
		6     & 3.726 & 7.780 & 1.557 & 9.401 \\
		7     & 3.727 & 7.778 & 2.007 & 8.694 \\
		8     & 3.727 & 7.778 & 2.550 & 8.254 \\
		9     & -     & -     & 3.050 & 8.025 \\
		10    & -     & -     & 3.379 & 7.924 \\
		11    & -     & -     & 3.539 & 7.884 \\
		12    & -     & -     & 3.603 & 7.869 \\
		13    & -     & -     & 3.630 & 7.863 \\
		14    & -     & -     & 3.634 & 7.862 \\
		15    & -     & -     & 3.638 & 7.861 \\
		16    & -     & -     & 3.639 & 7.861 \\
		17    & -     & -     & 3.639 & 7.861 \\
		\hline
	\end{tabular}%
	\label{iterations_SimulatedExamples}%
\end{table}%

\begin{table}[!h]
	\centering
	\caption{Weibull model parameters ($\beta$, $\eta$) and expected components' time to failure (${\rm E}(Y)$) estimation based on different estimation models: the Bayesian approach (BA), the EM maximum likelihood method (EM-ML) and the maximum likelihood approach from {\it Zhang et al.} (Z-ML) of simulated examples. There are no Z-ML estimates in Example 2 because it could not be computed due to the high number of components and failures.}
  \begin{threeparttable}
	\begin{tabular}{ccccccccc}
			\hline
		\multicolumn{9}{c}{BA} \\
			\hline
		&  \multicolumn{4}{c}{Example 1} & \multicolumn{4}{c}{Example 2} \\
			\hline
		Parameters & Mean  & SD    & \multicolumn{2}{c}{HPD 95\%} & Mean  & SD    & \multicolumn{2}{c}{HPD 95\%} \\
		\midrule
		$\beta$ & 3.696 & 0.321 & 3.095 & 4.357 & 3.638 & 0.118 & 3.410 & 3.858 \\
		$\eta$ & 7.890 & 0.522 & 6.906 & 8.912 & 7.861 & 0.074 & 7.733 & 8.007 \\
		$\rm{E}(Y)$   & 7.118 & 0.439 & 6.282 & 7.971 & 7.087 & 0.064 & 6.982 & 7.216 \\
			\hline
		\multicolumn{9}{c}{EM-ML} \\
			\hline
		&  \multicolumn{4}{c}{Example 1} & \multicolumn{4}{c}{Example 2} \\
			\hline
		Parameters & MLE   & SE    & \multicolumn{2}{c}{CI 95\%} & MLE   & SE    & \multicolumn{2}{c}{CI 95\%} \\
			\hline
		$\beta$ & 3.728 & 0.377 & 2.989 & 4.466 & 3.641 & 0.089 & 3.467 & 3.815 \\
	    $\eta$  & 7.777 & 0.585 & 6.631 & 8.924 & 7.860 & 0.053 & 7.757 & 7.964 \\
		$\rm{E}(Y)$   & 7.022 & 0.487 & 6.067 & 7.976 & 7.087 & 0.048 & 6.993 & 7.182 \\
			\hline
		\multicolumn{9}{c}{Z-ML} \\
			\hline
		&  \multicolumn{4}{c}{Example 1} & \multicolumn{4}{c}{Example 2} \\
			\hline
		Parameters & MLE   & SE    & \multicolumn{2}{c}{CI 95\%} & MLE   & SE    & \multicolumn{2}{c}{CI 95\%} \\
			\hline
		$\beta$ & 3.729 & 0.323 & 3.096 & 4.361 & -     & -     & -     & - \\
		$\eta$ & 7.776 & 0.488 & 6.820 & 8.732 & -     & -     & -     & - \\
		$\rm{E}(Y)$   & 7.021 & 0.409 & 6.218 & 7.823 & -     & -     & -     & - \\
		\hline
	\end{tabular}%
	\begin{tablenotes}
		\small
		\item SD means standard deviation; SE means standard error; HPD means highest posterior density; CI means confidence interval. The true parameters values are: $\beta=3.924$, $\eta=7.734$ and $\rm{E}(Y)=7$. 
	\end{tablenotes}
\end{threeparttable}
\label{tab_estParams_ExSim}%
\end{table}%

\begin{table}[!h]
	\centering
	\caption{MAE values obtained by the Bayesian approach (BA), the EM maximum likelihood method (EM-ML) and the maximum likelihood approach from {\it Zhang et al.} (Z-ML) of two simulated examples. There are no MAE values for Z-ML in Example 2 because they could not be computed due to the high number of components and failures.}
	\begin{tabular}{cccc}
		\hline
	       	& BA    & EM-ML & Z-ML \\
		\midrule
		Example 1 & 0.0117 & 0.0057 & 0.0057 \\
		Example 2  & 0.0080 & 0.0079 & - \\
		\hline
	\end{tabular}%
	\label{EMA_simulated_example}%
\end{table}%

\subsection{Simulation studies in different scenarios}

We conducted the simulations for all combinations of the following features: $n \in \{10, 50, 100, 200\}$, $m \in \{4, 8, 16, 32\}$, and $m_c \in \{4, 8\}$, resulting in $32$ scenarios.  For each scenario,  $100$ datasets were generated, and we compare the MAE from the estimators to the true distribution. 

\begin{figure}[h]\centering
 	\begin{minipage}[b]{0.47\linewidth}
 		\includegraphics[width=\linewidth]{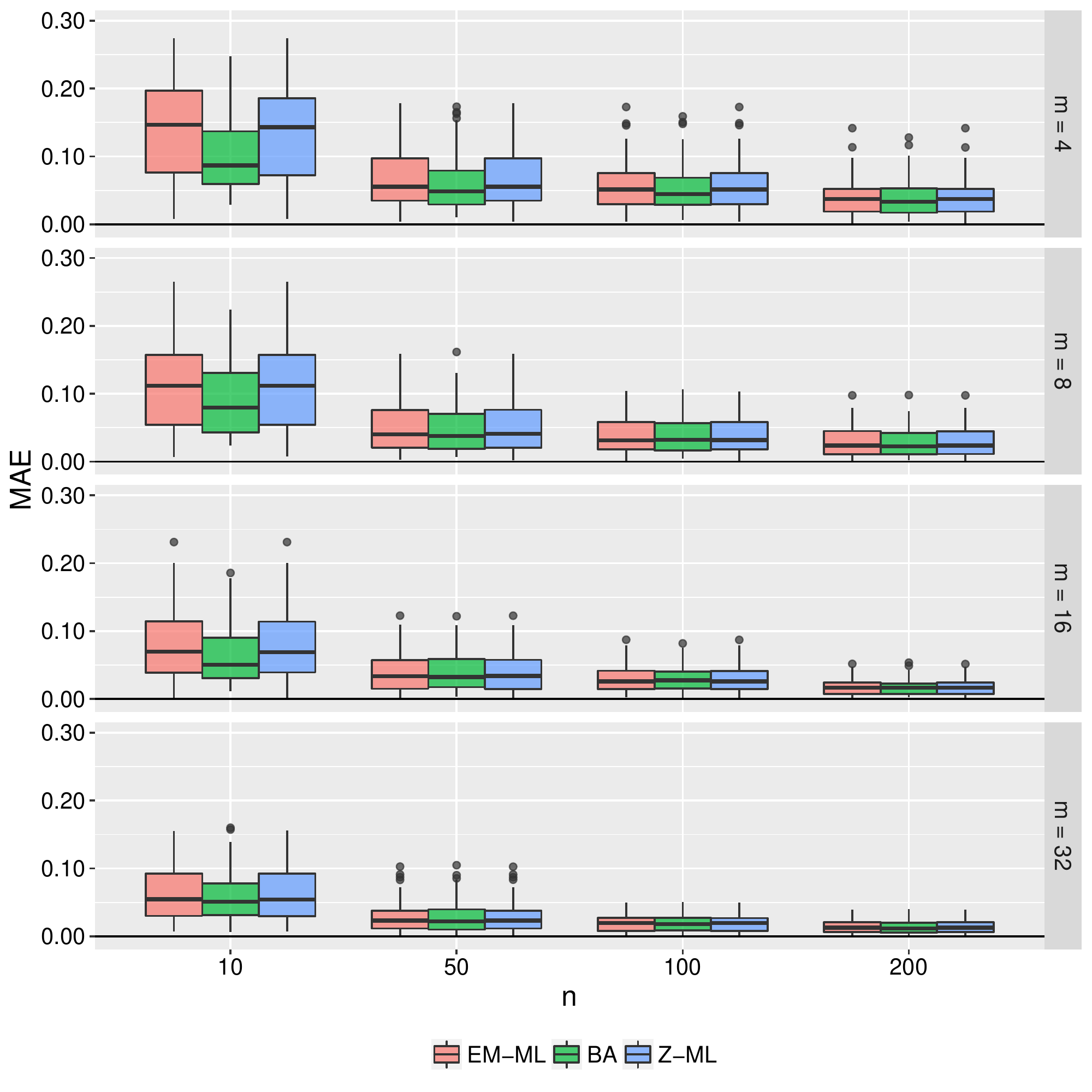}
 		\subcaption{ $\mbox{   }$ $m_c=4$}
 	\end{minipage} 
 	\begin{minipage}[b]{0.47\linewidth}
 		\includegraphics[width=\linewidth]{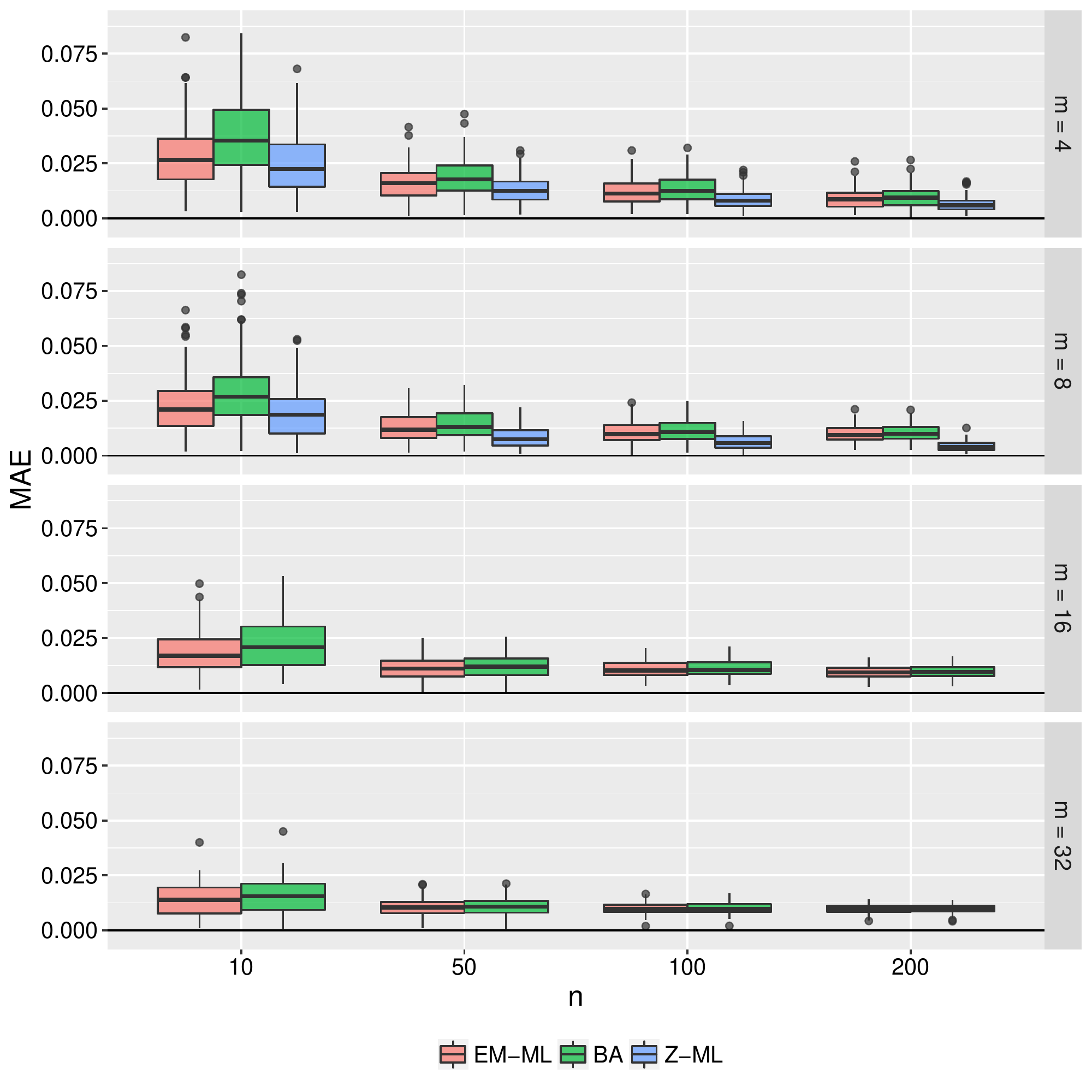}
 		\subcaption{ $\mbox{   }$ $m_c=8$}  \label{MAE_media4_media8_b}
 	\end{minipage}
 	\caption{Boxplot graphs of the $100$ MAE values of the Bayesian approach (BA), the EM maximum likelihood method (EM-ML) and the maximum likelihood approach from {\it Zhang et al.} (Z-ML) in scenarios with different sample sizes ($n$) and number of components ($m$). There are no Z-ML MAE boxplots in case of $m \in \{16, 32\}$ and $m_c = 8$ because they could not be computed due to the high number of components and failures.}
 	\label{MAE_media4_media8}
 \end{figure}

The boxplot graphs of $100$ MAE values are presented in $\mbox{Figure}~ \ref{MAE_media4_media8}$. In general, the methods present similar performance. When $m_c=8$ the BA method presents higher MAE means but the boxplot graph intersects with the boxplot graphs obtained by other methods. 

Noticeably, $\mbox{Figure}~\ref{MAE_media4_media8_b}$ does not contain any boxplots for Z-ML in case of $m \in \{16, 32\}$ and $m_c = 8$. However, this is plausible as this method was not able to compute the respective estimates due to the high number of failures and components. The computational time of each scenario was greater than four days and encountered errors in estimation. On the other hand, the computational times and availability of EM-ML and BA are not influenced that much by the numbers of failures and components.

In short, in settings as those from $\mbox{Figure}~\ref{MAE_media4_media8_b}$, Z-ML fails to compute the components' failure time distribution, whereas the two proposed methods find solutions. For the settings in which Z-ML finds solutions, the proposed methods also find solutions and present similar performance. 

\clearpage
 \section{Cylinder dataset analysis} \label{cylinder_data}

A fleet of $n=120$ diesel engines (systems) is observed. Each engine has $16$ identical cylinders working in series, that is, the first cylinder to fail causes the engine failure. When a cylinder fails, it is replaced by an identical functioning one in the socket (cylinder position), but the information about which socket each replacement comes from is not observed. $\mbox{Table}~\ref{dist_failures}$ presents the distributon of the number of failures across all 120 systems.

\begin{table}[!h]
	\centering
	\caption{Distributon of number of failures ($r$) of $120$ systems from cylinder dataset.}
	\begin{tabular}{c|c|c}
		\hline
		$r$ & Number of systems & \% \\
		\hline
		0     & 46    & 38.3 \\
		1     & 32    & 26.7 \\
		2     & 18    & 15.0 \\
		3     & 14    & 11.7 \\
		4     & 5     & 4.2 \\
		5     & 4     & 3.3 \\
		6     & 1     & 0.8 \\
		\hline
		Total & 120   & 100.0 \\
		\hline
	\end{tabular}%
	\label{dist_failures}%
\end{table}%

We fitted models assuming the following distributions for components' failure times: Weibull, gamma, lognormal and log-logistic. 
Under the frequentist approach, the lognormal model presents the lowest value for all selection criteria ($\mbox{Table}~\ref{aplic_freq}$) and as a consequence, it is the selected model. 

\begin{table}[htbp]
	\centering
	\caption{Selection criteria under frequentist approach obtained by the fitted models for cylinder dataset.}
	\begin{tabular}{c|cccccc}
		\toprule
		Model & $l$ & AIC   & AICc  & BIC   & HQIC  & CAIC \\
		\midrule
		Weibull & -677.81 & 1359.62 & 1359.72 & 1365.20 & 1361.89 & 1367.20 \\
		gamma & -673.86 & 1351.73 & 1351.83 & 1357.31 & 1353.99 & 1359.31 \\
		lognormal & -671.15 & 1346.30 & 1346.41 & 1351.88 & 1348.67 & 1353.88 \\
		log-logistic & -677.00 & 1357.99 & 1358.10 & 1363.57 & 1360.26 & 1365.57 \\
		\bottomrule
	\end{tabular}%
	\label{aplic_freq}%
\end{table}%

 Under the Bayesian paradigm, for each model, we run the Metropolis within Gibbs sampler, discarding the first $20,\!000$ as burn-in samples and using a jump of size $20$ to avoid correlation problems, obtaining a sample size of $1,\!000$. We evaluated the convergence of the chain by multiple runs of the algorithm from different starting values and the chains' convergence was monitored through graphical analysis, and good convergence results were obtained. Further, we considered the Gelman-Rubin convergence diagnostic statistics. The measures are close to $1$ for all parameters in all fitted models, as shown in $\mbox{Table}~\ref{aplic_bayes}$, which suggests that convergence chains have been reached.

The LPML values are presented in $\mbox{Table}~\ref{aplic_bayes}$ and the lognormal model is the chosen one once it presents the largest LPML value.

\begin{table}[htbp]
	\centering
	\caption{Gelman-Rubin Statistics and LPML measures obtained by the fitted models for cylinder dataset.}
	\begin{tabular}{c|cc}
		\toprule
		Model & Gelman-Rubin Statistics & LPML \\
		\hline
		Weibull & 1.0014 - 1.0024 & -687.05 \\
		gamma & 1.0032 -  1.0048 & -680.55 \\
		lognormal & 1.0020 - 1.0021 & -676.52 \\
		log-logistic & 1.0030  - 1.0033 & -687.08 \\
		\bottomrule
	\end{tabular}%
	\label{aplic_bayes}%
\end{table}%

$\mbox{Table}~ \ref{tab_estParams}$ lists the posterior mean obtained by BA and EM-ML estimates for the parameters of $\mu_l$ (mean of logarithm), $\sigma_l$ (standard deviation of logarithm) and expected time of components' lifetime, $\rm{E}(Y)=\exp\big\{\mu_l+\sigma_l^2/2\big\}$. 
The expected times of the component lifetime obtained by BA and EM-ML are $11.05$ and $10.93$ years, respectively. In general, the BA and EM-ML estimates are close for all parameters, as expected. 

\begin{figure}[!h]\centering
	\begin{minipage}[b]{0.47\linewidth}
		\includegraphics[width=\linewidth]{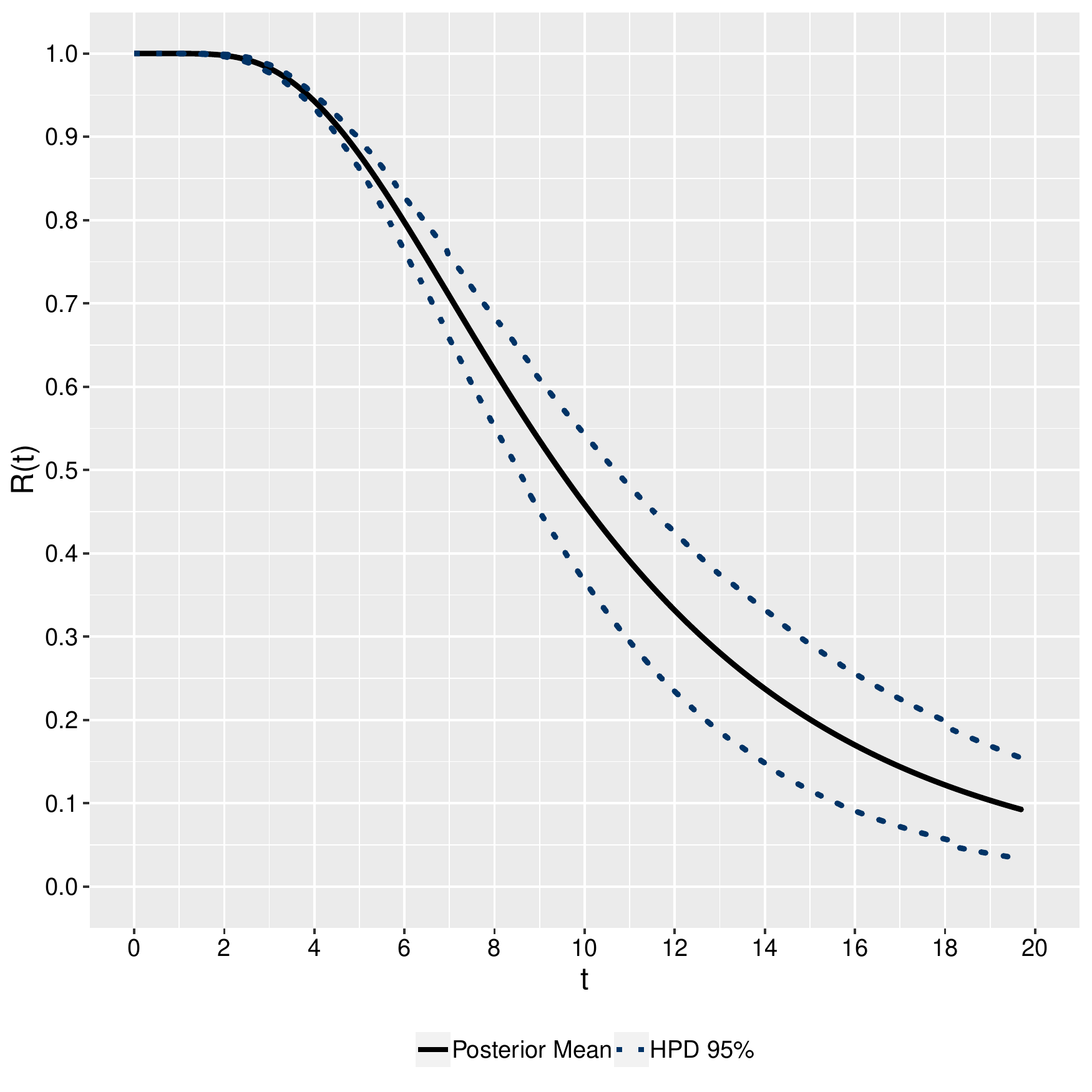}
		\subcaption{ $\mbox{   }$ Component reliability function}
		\label{fig:reliacylinder}
	\end{minipage} 
	\begin{minipage}[b]{0.47\linewidth}
		\includegraphics[width=\linewidth]{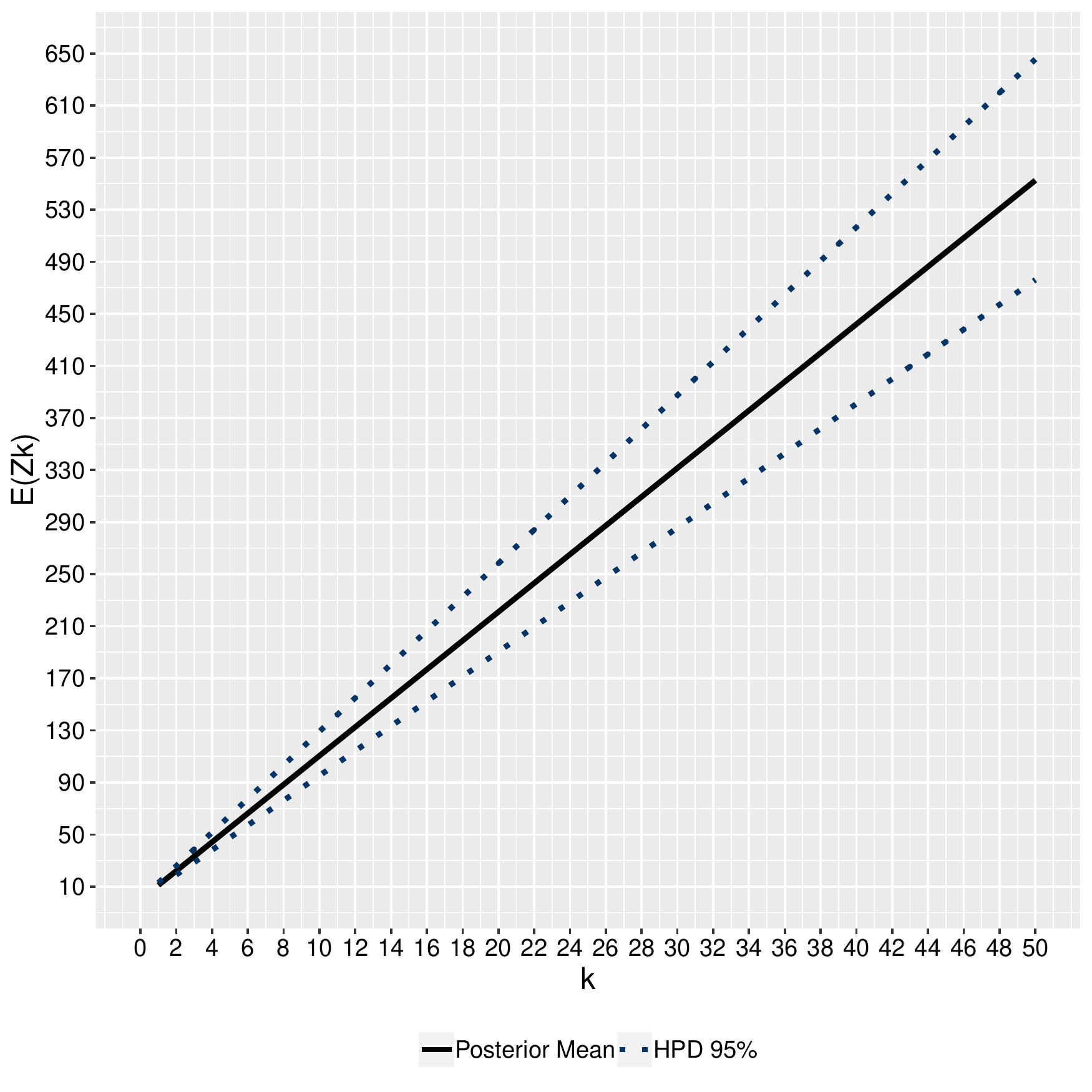}
		\subcaption{ $\mbox{   }$ $\rm{E}(Z_k)$ estimation}
		\label{fig:EZk_cylinder}
	\end{minipage}
	\caption{Component reliability function and expected time of occurence of the $k$-th failure in the socket estimates through Bayesian approach of cylinder dataset.}
	\label{cylinder_graph}
\end{figure}

The posterior mean and the $95\%$ highest posterior density (HPD) point-wise band of the component reliability function are illustrated in $\mbox{Figure}~ \ref{fig:reliacylinder}$. Besides, the posterior mean and the $95\%$ highest posterior density (HPD) point-wise band of ${\rm E}(Z_k)$, for $k=\{0,1,\ldots,49,50\}$, are presented in $\mbox{Figure}~ \ref{fig:EZk_cylinder}$. The estimation for the reliability function obtained by EM-ML estimator is similar to the estimate obtained by the Bayesian approach.

\begin{table}[!h]
	\centering
	\begin{threeparttable}
	 \caption{Lognormal model parameters ($\mu_l$, $\sigma_l$) and expected components' time to failure (${\rm E}(Y)$) estimation based on the Bayesian approach (BA) and the EM Maximum Likelihood method (EM-ML) of cylinder dataset.}
	\begin{tabular}{c|cccc|cccc}
		\hline
		& \multicolumn{4}{c|}{BA} & \multicolumn{4}{c}{EM-ML} \\
		\hline
		Parameters & Posterior Mean  & Posterior SD    & \multicolumn{2}{c|}{HPD 95\%} & MLE   & SE    & \multicolumn{2}{c}{CI 95\%} \\
		\hline
		$\mu_l$     &  2.2494 &  0.0597 &   2.1361 & 2.3677  & 2.2443  & 0.0952  & 2.0577  & 2.4309  \\
		$\sigma_l$  & 0.5464   & 0.0369  &  0.4749 & 0.6208  & 0.5433  &  0.0554 & 0.4346  & 0.6520  \\
		$\rm{E}(Y)$ & 11.0519  & 0.8887  &  9.5231 & 12.9137  &  10.9345       & 1.3673  & 8.2546   & 13.6144  \\
		\hline
	\end{tabular}%
    \label{tab_estParams}%
    \begin{tablenotes}
	\small
	\item SD means standard deviation; SE means standard error; HPD means highest posterior density and CI means confidence interval.
    \end{tablenotes}
   \end{threeparttable}
\end{table}%

\section{Conclusion} \label{final_remarks}

A Bayesian model and a maximum likelihood estimator (MLE) were proposed in order to estimate identical components failure time distribution involved in a repairable series system with masked cause of failure. For both approaches, latent variables were considered in the estimation process through EM algorithm for MLE and Markov-Chain Monte-Carlo (MCMC) for the Bayesian approach. 
The proposed models are generic and straightforward for any probability distribution on positive support. In estimation processes, satisfactory results about the convergence of the MCMC's chains and EM algorithm were obtained, evaluated through graphical analysis and convergence performance measures. 

Simulation studies were realized in scenarios with different sample sizes, number of components and distributions for censor lifetime. The mean absolute error (MAE) from each estimator to the true distribution was considered as performance measure. In situations of high numbers of failures and/or components, it was not possible to compute the maximum likelihood estimator proposed by \cite{zhang2017} (Z-ML) through the package \texttt{SRPML}. In contrast to this well-established approach by \cite{zhang2017}, our proposed methods are not affected by the high numbers of failures and/or components. Instead they work perfectly even in these situations. Besides, in settings in which Z-ML finds solutions, the proposed methods also find a solution and achieve a similar performance. Thus, the huge advantage of our proposed methods is that they estimate the components' failure time distribution regardless of the number of failures and components.
The practical applicability was assessed in cylinder dataset, in which components' failure time quantities were estimated convincingly.

In this work, the assumption of independent and identically distributed (i.i.d.) components failure times has been made and found to be suitable for the cylinder dataset characteristics. However, this assumption might not be applicable to other scenarios. Thus, in future works, our proposed method can be extended to situations in which the assumption of independent and identically distributed failure times is violated. Moreover, within future works we will also investigate the suitability of our approach for the assessment of system reliability rather than cylinder reliability, which has been the focus of this work.

\section*{Acknowledgment}

This work was partially supported by the Brazilian agency CNPq: grant 308776/2014-3. The agency had no role in the study design, data collection and analysis, decision to publish, or preparation of the manuscript.

This study was financed in part by CAPES (Brazil) - Finance Code 001 and Federal University of Mato Grosso do Sul.

Pascal Kerschke, Heike Trautmann, Bernd Hellingrath and Carolin Wagner acknowledge support by the European Research Center for Information Systems  (ERCIS).

\newpage 

\bibliographystyle{natbib}

\section*{Appendix}

We can write the logarithm of the augmented likelihood function of $i$-th system if Weibull distribution with parameter $\beta$ (shape) and $\eta$ (scale) is assumed, as
\begin{eqnarray} 
&&l_{i}(\bm{\theta}\mid\bm{t}_i,\bm{d}_i) = \Big[1-\rm{I}(v_i=0)\Big]\Bigg[\sum_{l=1}^{v_i}\sum_{k=1}^{n_l}\log f(x_{ilk}-x_{il(k-1)}) + \sum_{l=1}^{v_i} \log R(\tau_i-x_{iln_l}) \Bigg]+ (m-v_i)\log R(\tau_i) \nonumber \\
&=& \Big[1-\rm{I}(v_i=0)\Big]\sum_{l=1}^{v_i}\sum_{k=1}^{n_l}\Bigg\{\log(\beta)-\log(\eta)+(\beta-1)\Big[\log(x_{ilk}-x_{il(k-1)})-\log(\eta)\Big]-\Bigg(\frac{x_{ilk}-x_{il(k-1)}}{\eta}\Bigg)^{\beta}\Bigg\} \nonumber \\
&& - \Big[1-\rm{I}(v_i=0)\Big] \sum_{l=1}^{v_i} \Bigg(\frac{\tau_i-x_{iln_l}}{\eta}\Bigg)^{\beta} - (m-v_i)\Bigg(\frac{\tau_i}{\eta}\Bigg)^{\beta}. \nonumber \\
&=& \Big[1-\rm{I}(v_i=0)\Big]\Bigg\{r_i\log(\beta)-r_i\log(\eta)+(\beta-1)\sum_{l=1}^{v_i}\sum_{k=1}^{n_l}\log(x_{ilk}-x_{il(k-1)})-r_i(\beta-1)\log(\eta) \nonumber \\
&& -\sum_{l=1}^{v_i}\Bigg[\sum_{k=1}^{n_l}\Bigg(\frac{x_{ilk}-x_{il(k-1)}}{\eta}\Bigg)^{\beta} + \Bigg(\frac{\tau_i-x_{iln_l}}{\eta}\Bigg)^{\beta} \Bigg]\Bigg\} - (m-v_i)\Bigg(\frac{\tau_i}{\eta}\Bigg)^{\beta}. \nonumber 
\end{eqnarray}

The first derivatives 0f $l_{i}(\bm{\theta}\mid\bm{t}_i,\bm{d}_i)$ in relation to $\beta$ and $\eta$, respectively, are
\begin{eqnarray}
\frac{\d l_{i}(\bm{\theta}\mid\bm{t}_i,\bm{d}_i)}{\d\beta}= \Big[1-\rm{I}(v_i=0)\Big]\Bigg\{\frac{r_i}{\beta}+\sum_{l=1}^{v_i}\sum_{k=1}^{n_l}\log(x_{ilk}-x_{il(k-1)}) - r_i\log(\eta)+\log(\eta)\Bigg(\frac{1}{\eta}\Bigg)^{\beta}\Bigg[\sum_{l=1}^{v_i}\Bigg(\sum_{k=1}^{n_l} (x_{ilk}-x_{il(k-1)})^{\beta} \nonumber \\
+ (\tau_i-x_{iln_l})^{\beta}\Bigg) \Bigg]-\Bigg(\frac{1}{\eta}\Bigg)^{\beta}\Bigg[\sum_{l=1}^{v_i}\sum_{k=1}^{n_l}\log(x_{ilk}-x_{il(k-1)})(x_{ilk}-x_{il(k-1)})^{\beta}+\sum_{l=1}^{v_i}\log(\tau_i-x_{iln_l})(\tau_i-x_{iln_l})^{\beta}\Bigg]\Bigg\} \nonumber \\
+\Bigg(\frac{1}{\eta}\Bigg)^{\beta}(m-v_i)\tau_i^{\beta}[\log(\eta)-\log(\tau_i)], \nonumber
\end{eqnarray}
and
\begin{eqnarray}
\frac{\d l_{i}(\bm{\theta}\mid\bm{t}_i,\bm{d}_i)}{\d\eta}=\Big[1-\rm{I}(v_i=0)\Big]\Bigg\{-\frac{r_i}{\eta}-\frac{r_i(\beta-1)}{\eta}+\beta\Bigg(\frac{1}{\eta}\Bigg)^{\beta+1} \Bigg[\sum_{l=1}^{v_i}\sum_{k=1}^{n_l}(x_{ilk}-x_{il(k-1)})^{\beta}+\sum_{l=1}^{v_i}(\tau_i-x_{iln_l})^{\beta}\Bigg]\Bigg\} \nonumber \\
+\beta\Bigg(\frac{1}{\eta}\Bigg)^{\beta+1}(m-v_i)\tau_i^{\beta}. \nonumber
\end{eqnarray}

The second derivatives are
\begin{eqnarray}
\frac{\d^2 l_{i}(\bm{\theta}\mid\bm{t}_i,\bm{d}_i)}{\d\beta^2}=\Big[1-\rm{I}(v_i=0)\Big]\Bigg\{-\frac{r_i}{\beta^2} -[\log \eta]^2\Bigg(\frac{1}{\eta}\Bigg)^{\beta}\Bigg[\sum_{l=1}^{v_i}\Bigg(\sum_{k=1}^{n_l} (x_{ilk}-x_{il(k-1)})^{\beta}+ (\tau_i-x_{iln_l})^{\beta}\Bigg) \Bigg] \nonumber \\ +2\log(\eta)\Bigg(\frac{1}{\eta}\Bigg)^{\beta}\Bigg[\sum_{l=1}^{v_i}\sum_{k=1}^{n_l}\log(x_{ilk}-x_{il(k-1)})(x_{ilk}-x_{il(k-1)})^{\beta}+\sum_{l=1}^{v_i}\log(\tau_i-x_{iln_l})(\tau_i-x_{iln_l})^{\beta}\Bigg] \nonumber \\
-\Bigg(\frac{1}{\eta}\Bigg)^{\beta}\Bigg[\sum_{l=1}^{v_i}\sum_{k=1}^{n_l}[\log(x_{ilk}-x_{il(k-1)})]^2(x_{ilk}-x_{il(k-1)})^{\beta}+\sum_{l=1}^{v_i}[\log(\tau_i-x_{iln_l})]^2(\tau_i-x_{iln_l})^{\beta}\Bigg] \Bigg\} \nonumber \\
+(m-v_i)\Bigg(\frac{1}{\eta}\Bigg)^{\beta}\tau_i^{\beta}\Bigg[-[\log(\tau_i)]^2 +2\log(\tau_i)\log(\eta)-[\log(\eta)]^2\Bigg], \nonumber
\end{eqnarray} 
\begin{eqnarray}
\frac{\d^2 l_{i}(\bm{\theta}\mid\bm{t}_i,\bm{d}_i)}{\d\beta\d\eta}=\Big[1-\rm{I}(v_i=0)\Big]\Bigg\{-\frac{r_i}{\eta} + \Bigg[\Bigg(\frac{1}{\eta}\Bigg)^{\beta+1}(1-\beta\log(\eta))\Bigg]\Bigg[\sum_{l=1}^{v_i}\Bigg(\sum_{k=1}^{n_l} (x_{ilk}-x_{il(k-1)})^{\beta}+ (\tau_i-x_{iln_l})^{\beta}\Bigg) \Bigg] \nonumber \\
+ \beta\Bigg(\frac{1}{\eta}\Bigg)^{\beta+1}\Bigg[\sum_{l=1}^{v_i}\sum_{k=1}^{n_l}\log(x_{ilk}-x_{il(k-1)})(x_{ilk}-x_{il(k-1)})^{\beta}+\sum_{l=1}^{v_i}\log(\tau_i-x_{iln_l})(\tau_i-x_{iln_l})^{\beta}\Bigg] \Bigg\} \nonumber \\
+(m-v_i)\Bigg(\frac{1}{\eta}\Bigg)^{\beta+1}\tau_i^{\beta}[1-\beta\log(\eta)+\beta\log(\tau_i)], \nonumber 
\end{eqnarray}
and 
\begin{eqnarray}
\frac{\d^2  l_{i}(\bm{\theta}\mid\bm{t}_i,\bm{d}_i)}{\d\eta^2 }=\Big[1-\rm{I}(v_i=0)\Big]\Bigg\{\frac{\beta r_i}{\eta^2}-\beta(\beta+1)\Bigg(\frac{1}{\eta}\Bigg)^{\beta+2} \Bigg[\sum_{l=1}^{v_i}\Bigg(\sum_{k=1}^{n_l} (x_{ilk}-x_{il(k-1)})^{\beta}+ (\tau_i-x_{iln_l})^{\beta}\Bigg) \Bigg]\Bigg\}  \nonumber \\
-\beta(\beta+1)\Bigg(\frac{1}{\eta}\Bigg)^{\beta+2}(m-v_i)\tau_i^{\beta}. \nonumber 
\end{eqnarray}

Thus,
\begin{eqnarray}
I=-\frac{\partial^2}{\partial\bm{\theta}\partial\bm{\theta}^\top}Q(\bm{\theta}\mid\widehat{\bm{\theta}})=-\frac{1}{L}\sum_{i=1}^n\sum_{l=1}^L\frac{\partial^2}{\partial\bm{\theta}\partial\bm{\theta}^\top}l_{i}\Big(\bm{\theta}\mid\bm{t}_i,\bm{d}_{i}^{(l)}\Big)\Bigg|_{\bm{\theta}=\widehat{\bm{\theta}}}= \nonumber 
\end{eqnarray}
\[
\left[ {\begin{array}{cc}
	-\frac{1}{L}\sum_{i=1}^n\sum_{l=1}^L \frac{\d^2  l_{i}(\bm{\theta}\mid\bm{t}_i,\bm{d}_{i}^{(l)})}{\d\eta^2 }\Bigg|_{\bm{\theta}=\widehat{\bm{\theta}}} & 	-\frac{1}{L}\sum_{i=1}^n\sum_{l=1}^L \frac{\d^2  l_{i}(\bm{\theta}\mid\bm{t}_i,\bm{d}_{i}^{(l)})}{\d\eta\d\beta }\Bigg|_{\bm{\theta}=\widehat{\bm{\theta}}}  \\
	-\frac{1}{L}\sum_{i=1}^n\sum_{l=1}^L \frac{\d^2  l_{i}(\bm{\theta}\mid\bm{t}_i,\bm{d}_{i}^{(l)})}{\d\beta\d\eta }\Bigg|_{\bm{\theta}=\widehat{\bm{\theta}}}  & -\frac{1}{L}\sum_{i=1}^n\sum_{l=1}^L \frac{\d^2  l_{i}(\bm{\theta}\mid\bm{t}_i,\bm{d}_{i}^{(l)})}{\d\beta^2 }\Bigg|_{\bm{\theta}=\widehat{\bm{\theta}}} 
	\end{array} } \right], 
\]\nonumber
in which $\widehat{\bm{\theta}}=(\widehat{\eta},\widehat{\beta})$.
Besides, 
\begin{eqnarray}
&&II=\sum_{i=1}^n\Bigg\{\frac{1}{L}\sum_{l=1}^L\frac{\partial}{\partial\bm{\theta}}l_{i}\Big(\bm{\theta}\mid\bm{t}_i,\bm{d}_{i}^{(l)}\Big)\Bigg|_{\bm{\theta}=\widehat{\bm{\theta}}}\Bigg\}\Bigg\{\frac{1}{L}\sum_{l=1}^L\frac{\partial}{\partial\bm{\theta}}l_{i}\Big(\bm{\theta}\mid\bm{t}_i,\bm{d}_{i}^{(l)}\Big)\Bigg|_{\bm{\theta}=\widehat{\bm{\theta}}}\Bigg\}^\top \nonumber \\
&&=\sum_{i=1}^n\Bigg\{\frac{1}{L}\sum_{l=1}^L\Bigg( \frac{\d  l_{i}(\bm{\theta}\mid\bm{t}_i,\bm{d}_{i}^{(l)})}{\d\eta }\Bigg|_{\bm{\theta}=\widehat{\bm{\theta}}},  \frac{\d  l_{i}(\bm{\theta}\mid\bm{t}_i,\bm{d}_{i}^{(l)})}{\d\beta }\Bigg|_{\bm{\theta}=\widehat{\bm{\theta}}}\Bigg)^\top\Bigg\}\Bigg\{\frac{1}{L}\sum_{l=1}^L\Bigg( \frac{\d  l_{i}(\bm{\theta}\mid\bm{t}_i,\bm{d}_{i}^{(l)})}{\d\eta }\Bigg|_{\bm{\theta}=\widehat{\bm{\theta}}},  \frac{\d  l_{i}(\bm{\theta}\mid\bm{t}_i,\bm{d}_{i}^{(l)})}{\d\beta }\Bigg|_{\bm{\theta}=\widehat{\bm{\theta}}}\Bigg)^\top\Bigg\}^{\top} \nonumber
\end{eqnarray}
and
\begin{eqnarray}
&& III= -\frac{1}{L}\sum_{i=1}^n\sum_{l=1}^L\Bigg\{\frac{\partial}{\partial\bm{\theta}}l_{i}\Big(\bm{\theta}\mid\bm{t}_i,\bm{d}_{i}^{(l)}\Big)\Bigg\}\Bigg\{\frac{\partial}{\partial\bm{\theta}}l_{i}\Big(\bm{\theta}\mid\bm{t}_i,\bm{d}_{i}^{(l)}\Big)\Bigg\}^\top\Bigg|_{\bm{\theta}=\widehat{\bm{\theta}}} \nonumber \\
&& =-\frac{1}{L}\sum_{i=1}^n\sum_{l=1}^L\Bigg\{\Bigg( \frac{\d  l_{i}(\bm{\theta}\mid\bm{t}_i,\bm{d}_{i}^{(l)})}{\d\eta },  \frac{\d  l_{i}(\bm{\theta}\mid\bm{t}_i,\bm{d}_{i}^{(l)})}{\d\beta }\Bigg)^\top\Bigg\}\Bigg\{\Bigg( \frac{\d  l_{i}(\bm{\theta}\mid\bm{t}_i,\bm{d}_{i}^{(l)})}{\d\eta },  \frac{\d  l_{i}(\bm{\theta}\mid\bm{t}_i,\bm{d}_{i}^{(l)})}{\d\beta }\Bigg)^\top\Bigg\}^\top\Bigg|_{\bm{\theta}=\widehat{\bm{\theta}}}. \nonumber
\end{eqnarray}

The quantity $I_{\bm{\theta}}(\widehat{\bm{\theta}})$ can be estimated by $I+II+III$. 
\clearpage

\end{document}